# Superluminal Electromagnetic and Gravitational Fields Generated in the Nearfield of Dipole Sources

**William D. Walker**
Norwegian University of Science and Technology (NTNU)
Previous Research papers [1]
william.walker@vm.ntnu.no

## Abstract

In this paper the fields generated by an electric dipole and a gravitational quadrapole are shown to propagate superluminally in the nearfield of the source and reduce to the speed of light as the fields propagate into the farfield. A theoretical derivation of the generated fields using Maxwell's equations is presented followed by a theoretical analysis of the phase and group speed of the propagating fields. This theoretical prediction is then verified by a numerical simulation which demonstrates the superluminal propagation of modulated signals in the nearfield of their sources. An experiment using simple dipole antennas is also presented which verifies the theoretically expected superluminal propagation of transverse electromagnetic fields in the nearfield of the source. The phase speed, group speed, and information speed of these systems are compared and shown to differ. Provided the noise of a signal is small and the modulation method is known, it is shown that the information speed can be approximately the same as the superluminal group speed. According to relativity theory, it is known that between moving reference frames, superluminal signals can propagate backwards in time enabling violations of causality. Several explanations are presented which may resolve this dilemma.

## Introduction

The electromagnetic fields generated by an oscillating electric dipole have been theoretically studied by many researchers using Maxwell's equations. Typical analysis involves using potentials and an arbitrary gauge equation which simplifies the resultant PDEs so that they can be simply solved. In the following section, an analysis of the electric dipole is presented which solves the generated fields without the use of potentials and a gauge equation. The results show the same field solutions as presented by other authors but differ in that the electric field is shown to be generated only by the position of the dipole and not the combination of the dipole's position, velocity and acceleration. Similarly the magnetic field is shown to be generated only by the velocity of the dipole and not the combination of the dipole's, velocity and acceleration. Although the analysis presented below does not use potentials, the results are grouped in terms of potentials so that the results can be easily compared to the analysis presented by other authors. It should be noted that this paper is a summary of ongoing research done by the author since 1990 [ref. previous authors papers: 2, 3, 4, 5, 6, 7, 8, 9]. The analysis presented below is in (Gaussian units).

**Dipole field analysis**

**Fig. 1**

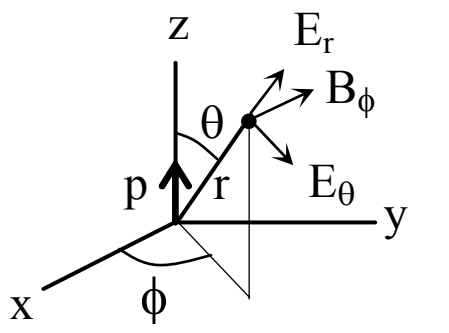

Variable definitions

$E_r$ = Radial electric field
$E_\theta$ = Transverse electric field
$B_\phi$ = Transverse magnetic field
V = Scalar potential
A = Vector potential
ρ = Charge density
J = Current density
$\varepsilon_o$ = Free-space permittivity
c = Speed of light
t = Time
p = Dipole
ω = Angular frequency
k = Wave number

The following derivation of the electromagnetic fields generated by a dipole source will be use the following well known free space Maxwell's equations:

Free space Maxwell equations (Gaussian units)

$$\nabla \cdot E = 4\pi\rho \quad \text{(Gauss's law for E field)} \tag{1}$$

$$\nabla \cdot B = 0 \quad \text{(Gauss's law for B field)} \tag{2}$$

$$\nabla \times E + \frac{1}{c}\frac{\partial B}{\partial t} = 0 \quad \text{(Faraday's law)} \tag{3}$$

$$\nabla \times B - \frac{1}{c}\frac{\partial E}{\partial t} = \frac{4\pi}{c}J \quad \text{(Ampère's law)} \tag{4}$$

Taking the curl of Faradays's law (Eq. 3) yields:

$$\nabla \times (\nabla \times E) + \frac{1}{c}\left(\nabla \times \frac{\partial B}{\partial t}\right) = 0 \tag{5}$$

Substituting the vector identity:

$$\nabla \times (\nabla \times E) = \nabla(\nabla \cdot E) - \nabla^2 E \tag{6}$$

Equation (5) becomes:

$$\nabla(\nabla \cdot E) - \nabla^2 E + \frac{1}{c}\frac{\partial}{\partial t}(\nabla \times B) = 0 \tag{7}$$

Using Gauss's law for E field (Eq. 1) and Ampère's law (Eq. 4), the above equation (Eq. 7) becomes:

$$\nabla(4\pi\rho) - \nabla^2 E + \frac{1}{c}\frac{\partial}{\partial t}\left(\frac{1}{c}\frac{\partial E}{\partial t} + \frac{4\pi}{c}J\right) = 0 \tag{8}$$

Reorganizing the above equation (8) yields the following 2nd order PDE for the E field:

$$\boxed{\nabla^2 E - \frac{1}{c^2}\frac{\partial^2 E}{\partial t^2} = 4\pi\nabla\rho + \frac{4\pi}{c^2}\frac{\partial J}{\partial t}} \tag{9}$$

A 2nd order PDE for the B field can be determined in a similar manor by taking the curl of Ampère's law (Eq. 4):

$$\nabla\times(\nabla\times B)-\frac{1}{c}\left(\nabla\times\frac{\partial E}{\partial t}\right)=\frac{4\pi}{c}\left(\nabla\times J\right) \tag{10}$$

Substituting the vector identity for B (ref. Eq. 6) yields:

$$\nabla(\nabla\cdot B)-\nabla^2 B-\frac{1}{c}\frac{\partial}{\partial t}\left(\nabla\times E\right)=\frac{4\pi}{c}\left(\nabla\times J\right) \tag{11}$$

Inserting Gauss's law for B field (Eq. 2) and Faraday's law (Eq. 3) in the above equation (Eq. 11) yields the following 2$^{nd}$ order PDE for the B field:

$$\boxed{\nabla^2 B-\frac{1}{c^2}\frac{\partial^2 B}{\partial t^2}=\frac{-4\pi}{c}\left(\nabla\times J\right)} \tag{12}$$

To solve these PDEs, assume they are of the following general form (ref. Ch 6.6 [10]):

$$\nabla^2\psi(\overline{R},t)-\frac{1}{c^2}\frac{\partial^2\psi(\overline{R},t)}{\partial t^2}=f(\overline{R},t) \tag{13}$$

where ($\overline{R}=\overline{r}-\overline{r}'$) is the distance from the charge to the observation point (Fig 2).

**Fig. 2**

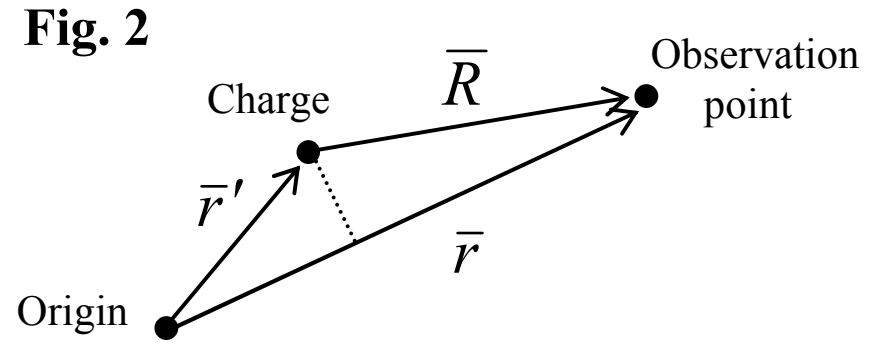

The Green's function method can then be used to solve this PDE equation. This is done by setting source f(t) to a Dirac delta function and solving for the Green's function (G) yielding:

$$\nabla^2 G(\overline{R},t)-\frac{1}{c^2}\frac{\partial^2 G(\overline{R},t)}{\partial t^2}=\delta(\overline{r}-\overline{r}')\delta(t-t') \tag{14}$$

The resulting equation can be simplified using the following known Fourier Transform relations:

$$\psi(\omega)=\int_{-\infty}^{\infty}\psi(t)e^{i\omega t}dt \qquad \text{Fourier Transform} \tag{15}$$

$$\psi(t)=\frac{1}{2\pi}\int_{-\infty}^{\infty}\psi(\omega)e^{-i\omega t}d\omega \qquad \text{Inverse Fourier Transform} \tag{16}$$

The PDE (Eq. 14) can then be simplified by inserting (Eq. 16) for $G(R,t)$ in the 2nd term and differentiating twice with respect to time yielding:

$$\nabla^2 G(\overline{R},t)-\frac{1}{c^2}(-i\omega)^2 G(\overline{R},t)=\delta(\overline{r}-\overline{r}')\delta(t-t') \tag{17}$$

This equation can then be multiplied by: $e^{i\omega t}$ and then integrated with respect to time. The resultant equation can then be simplified using the Fourier Transform (Eq. 15) and using the known property: $\int f(x)\delta(x-a)dx = f(a)$ yielding:

$$\nabla^2 G(\overline{R},\omega) + \left(\frac{\omega}{c}\right)^2 G(\overline{R},\omega) = \delta(\overline{r}-\overline{r}')e^{i\omega t'} \tag{18}$$

Since the Dirac delta function is zero anywhere outside the source, the PDE then reduces to the following form:

$$\nabla^2 G_k(\overline{R},\omega) + \left(\frac{\omega}{c}\right)^2 G_k(\overline{R},\omega) = 0 \tag{19}$$

Inserting the known relation for the Laplacian operator in spherical coordinates and assuming spherical symmetry in the resultant propagating fields (i.e. $G$ independent of $\theta$ and $\phi$) yields:

$$\frac{\partial}{\partial R^2}\left[RG_k(R,\omega)\right] + \left(\frac{\omega}{c}\right)^2 \left[RG_k(R,\omega)\right] = 0 \qquad \text{where } R = \left|\overline{R}\right| \tag{20}$$

The solution of this differential equation is known to be:

$$G_k(R,\omega) = \frac{Ae^{i\frac{\omega}{c}R} + Be^{-i\frac{\omega}{c}R}}{R} \tag{21}$$

Values for unknowns (A and B) can then be determined using known boundary conditions. As $\omega \to 0$ then governing differential equation (Eq. 18) reduces to Poisson's equation and it's known solution (ref. Ch. 1.7 [10]), implying: $G_k(R,0) = -1/(4\pi R)$, yielding: A+B=-1/(4$\pi$). Using the radiation condition, where it is assumed that the fields only propagate away from the source yields: B=0. Consequently the resultant solution for (Eq. 19) is:

$$G_k(R,\omega) = \frac{-e^{i\frac{\omega}{c}R}}{4\pi R} \tag{22}$$

The Green's function can then be determined by convoluting the above solution with the right side of the governing differential equation (Eq. 18), where the convolution relation is known to be: $y(R) = \int_{-\infty}^{\infty} x(R')h(R-R')dR'$ and then simplified using the known property: $\int f(x)\delta(x-a)dx = f(a)$ yielding:

$$G(R,\omega) = \int_{-\infty}^{\infty} \delta(R-R')e^{i\omega t'}G_k(R')dR' = G_k(R,\omega)e^{i\omega t'} \tag{23}$$

Using the inverse Fourier Transform relation (Eq. 16) yields:

$$G(R,t) = \frac{-1}{4\pi}\frac{1}{2\pi}\int_{-\infty}^{\infty} \frac{e^{ikR}}{R} e^{-i\omega(t-t')}d\omega \qquad \text{where: } k \equiv \frac{\omega}{c} \tag{24}$$

Using the following known relation for the Dirac delta function [11]:

$$\delta(x-y)=\frac{1}{2\pi}\int_{-\infty}^{\infty}e^{i\omega(x-y)}d\omega \tag{25}$$

The Green's function reduces to:

$$G=\frac{-1}{4\pi}\frac{\delta(t'-[t-R/c])}{R} \tag{26}$$

The fields (E, B) can then be determined by convoluting (*) the above Green's function (Eq. 26) with the RHS of the PDEs (Eq. 9, 12):

$$E=G*\left(4\pi\nabla\rho(t)+\frac{4\pi}{c^2}\frac{\partial J(t)}{\partial t}\right)\quad=\nabla\left(4\pi[G*\rho(t)]\right)+\frac{4\pi}{c^2}\frac{\partial[G*J(t)]}{\partial t} \tag{27}$$

$$B=G*\left(\frac{-4\pi}{c}(\nabla\times J(t))\right)\quad=\nabla\times\left(\frac{-4\pi}{c}[G*J(t)]\right) \tag{28}$$

where the convolution relation (*) is: $y=\int\int h(R-R',t-t')x(R',t')dt'dv'$. Because they correspond to different spatial coordinates, the $G(R',t')$ can be placed inside the gradient and curl which are functions of (R).

$$E=\nabla\left\{4\pi\int\left[\int_{-\infty}^{\infty}G\rho(t')dt'\right]dv'\right\}+\frac{1}{c}\frac{\partial}{\partial t}\left\{\frac{4\pi}{c}\int\left[\int_{-\infty}^{\infty}GJ(t')dt'\right]dv'\right\} \tag{29}$$

$$B=\nabla\times\left\{\frac{-4\pi}{c}\int\left[\int_{-\infty}^{\infty}GJ(t')dt'\right]dv'\right\} \tag{30}$$

These field equations can then be put in a more familiar form:

$$E=-\nabla V-\frac{1}{c}\frac{\partial A}{\partial t}\qquad B=\nabla\times A \tag{31}$$

where: $$V=-4\pi\int\left[\int_{-\infty}^{\infty}G\rho(t')\,dt'\right]dv'\qquad A=\frac{-4\pi}{c}\int\left[\int_{-\infty}^{\infty}GJ(t')\,dt'\right]dv' \tag{32}$$

Inserting the resultant Green's function (Eq. 26) yields:

$$V=\int\left[\frac{1}{R}\int_{-\infty}^{\infty}\delta(t'-[t-R/c])\rho(t')dt'\right]dv' \tag{33}$$

$$A=\frac{1}{c}\int\left[\frac{1}{R}\int_{-\infty}^{\infty}\delta(t'-[t-R/c])J(t')dt'\right]dv' \tag{34}$$

These relations can then be simplified by using the property: $\int f(x)\delta(x-a)dx = f(a)$ yielding:

$$V = \int \frac{\rho(t-\frac{R}{c})}{R} dv' \qquad A = \frac{1}{c}\int \frac{J(t-\frac{R}{c})}{R} dv' \tag{35}$$

The above potentials are commonly referred to in the literature as retarded potentials, (ref. p.681 [12]). Note that although the fields (E, B) have been derived without the use of the scalar potential (V) and vector potential (A), their solutions have been grouped using them, enabling comparison with other traditional derivations. Note that avoiding the use of potentials in the derivation of the fields eliminates any confusion involved in using arbitrary gauge equations used to solve potential PDEs [2].

For an oscillating signal, the sources become:

$$\rho(r') = \rho_o(r')e^{-i\omega t} \qquad J(r') = J_o(r')e^{-i\omega t} \tag{36}$$

Inserting these sources into the above potentials (Eq. 35) yields:

$$V = \int \rho(r')\frac{e^{ikR}}{R} dv' \qquad A = \frac{1}{c}\int J(r')\frac{e^{ikR}}{R} dv' \tag{37}$$

$$\text{where } R = |\bar{r} - \bar{r}'| \qquad k \equiv \frac{\omega}{c}$$

To simplify the above potential integrals (Eq. 37), the $\frac{f(kR)}{R}$ term can be Taylor series expanded about ($\xi = {r'}/{r}$) yielding:

**Fig. 3**

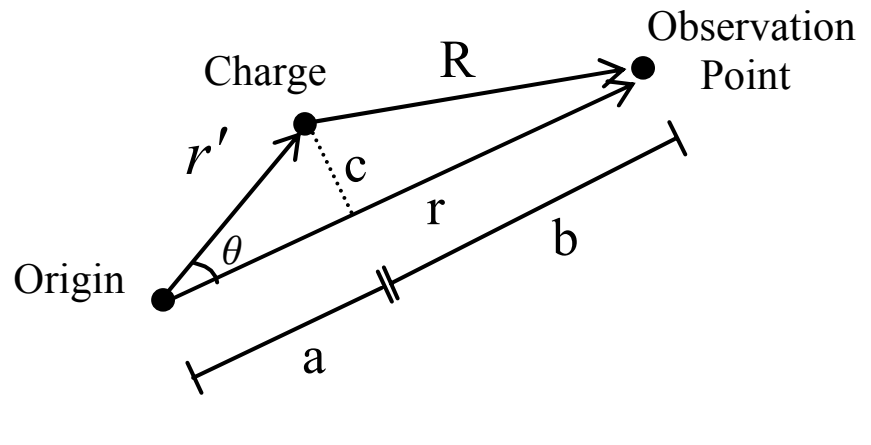

$$R = \sqrt{b^2 + c^2} \qquad r = a + b$$
$$a = r_o Cos(\theta) \qquad c = r_o Sin(\theta)$$

$$\therefore R = \sqrt{[r - r'Cos(\theta)]^2 + [r'Sin(\theta)]^2} \tag{38}$$

$$\frac{f(kR)}{R} = \frac{1}{r}f(kr) + \left[\frac{1}{r}f(kr) - kf'(kr)\right]\xi Cos(\theta) + O(\xi)^2 \tag{39}$$

$$\text{where } f'(kr) = \frac{\partial f(u)}{\partial u} = \frac{\partial f(u)}{\partial r}\frac{\partial r}{\partial u} = ie^{ikr}$$

note: $u = kr$, $f(kr) = e^{ikr}$, $\xi = \frac{r'}{r}$

**Fig. 4** Maple 7 code for Taylor series expansion:

```
R:=r*((1-a*cos(theta))^2+(a*sin(theta))^2)^(1/2);
series(f(k*R-w*t)/R,a=0,2);
```

$$\therefore \frac{e^{ikr}}{R} = \frac{e^{ikr}}{r} + \frac{e^{ikr}}{r}[1 - ikr]\xi Cos(\theta) + O(\xi)^2 \tag{40}$$

where $\xi = \frac{r'}{r}$, which is small if the observation point (r) is much farther than the dipole length ($r'$)

To zeroth order in ($\xi = \frac{r'}{r}$) the potentials (ref. Eqs. 37) are:

$$V = \frac{e^{ikr}}{r}\int \rho(r')dv' + O(\xi) \qquad A = \frac{1}{c}\frac{e^{ikr}}{r}\int J(r')dv' + O(\xi) \tag{41}$$

Given:

$$\int \rho(r')dv' = q = \frac{p}{r'} \qquad \int J(r')dv' = r'\frac{dq}{dt} = \frac{dp}{dt} = -i\omega p_o e^{-i\omega t} \tag{42}$$

where: $p = p_o e^{-i\omega t}$

yields:

$$\boxed{\therefore V = \frac{p}{r'}\frac{e^{ikr}}{r} + O(\xi)} \qquad \boxed{\therefore A = \frac{1}{c}\frac{dp}{dt}\frac{e^{ikr}}{r} + O(\xi)} \tag{43}$$

The B field can then be determined using equation (ref. Eq. 31): $B = \nabla \times A$ (44)

If the source is a vertical dipole p(z):

$$p(z) = p\{Cos(\theta)\hat{r} - Sin(\theta)\hat{\theta}\} \tag{45}$$

the B field becomes:

**Fig. 5**

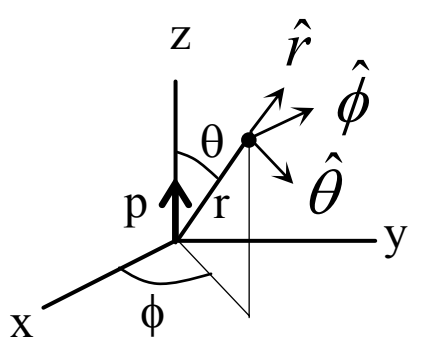

**Fig. 6** Maple 7 code for derivation of ($B_\phi$ $E_r$ $E_\theta$) from A:

```
restart;
with(linalg):
H := [r, theta, phi];
R:=r*((1-a*cos(theta))^2+(a*sin(theta))^2)^(1/2);
series(exp(I*k*R)/R,a=0,2);
s:=exp(I*k*r)/r/c*diff(p(t),t);
A:=[s*cos(theta),-s*sin(theta),0];
B:=curl(A,H,coords=spherical);
Bphi:=simplify(B[3]);
E:=curl(B,H,coords=spherical);
Er:=simplify(expand(c*(int(E[1],t))));
Eth:=simplify(expand(c*(int(E[2],t))));
p(t):=Po*exp(-I*w*t);
Bhi:=simplify(Bphi);
Er:=simplify(Er);
Eth:=simplify(Eth);
```

$$B_\phi = \frac{-Sin(\theta)}{cr^2}\left[\frac{dp}{dt}\right]e^{ikr}[ikr - 1] + O(\xi) \tag{46}$$

Inserting $p = p_o e^{-i\omega t}$ yields the B field to zeroth order in ($\xi = \frac{r'}{r}$):

$$\boxed{B_\phi = \frac{\omega p_o Sin(\theta)}{cr^2} e^{i(kr-\omega t)}\left[-kr - i\right] + O(\xi)} \tag{47}$$

Although the E field can be calculated using the potentials (Eq. 31): $E = -\nabla V - \frac{1}{c}\frac{\partial A}{\partial t}$, it is easier to use Ampère's law (Eq. 4): $\nabla \times B - \frac{1}{c}\frac{\partial E}{\partial t} = \frac{4\pi}{c} J$, which when solved for E field yields:

$$E = -4\pi \int J dt + c\int (\nabla \times B) dt \tag{48}$$

The first integral term is zero since $J = J_o e^{-i\omega t}$ is oscillatory. Inserting $B_\phi$ (Eq. 47) in the above equation and performing the time integral of the curl in spherical coordinates (ref. Fig. 6) yields:

$$E = E_r + E_\theta \tag{49}$$

where:

$$E_r = \frac{2Cos(\theta)}{r^3}[p]e^{ikr}\left[1 - i(kr)\right] + O(\xi) \tag{50}$$

$$E_\theta = \frac{Sin(\theta)}{r^3}[p]e^{ikr}\left[\left\{1 - (kr)^2\right\} - i(kr)\right] + O(\xi) \tag{51}$$

Inserting $p = p_o e^{-i\omega t}$ yields the E fields to zeroth order in ($\xi = \frac{r'}{r}$):

$$\boxed{E_r = \frac{2p_o Cos(\theta)}{r^3} e^{i(kr-\omega t)}\left[1 - i(kr)\right] + O(\xi)} \tag{52}$$

$$\boxed{E_\theta = \frac{p_o Sin(\theta)}{r^3} e^{i(kr-\omega t)}\left[\left\{1 - (kr)^2\right\} - i(kr)\right] + O(\xi)} \tag{53}$$

The above field solutions (Eq. 47, 52, 53) are the same as those presented by other authors [29, 12(ch 37.6)] but it is noted that they differ in that the magnetic field is shown to be generated only by the velocity of the dipole (Eq. 46) and not the combination of the dipole's, velocity and acceleration. Similarly the electric field is shown to be generated only by the position of the dipole (Eq. 50, 51) and not the combination of the dipole's position, velocity and acceleration. The above analysis shows that the real and imaginary components in the brackets (Eq. 47, 52, 53) causing the fields to deviate from the speed of light result from the Taylor expansion of the spatial part of the fields (Eq. 40) and not due to time derivatives of the source as suggested by other authors.

**Phase analysis**

The general form of the electromagnetic fields generated by a dipole is:

$$Field \propto (x + iy) \cdot e^{i[kr - \omega t]}$$

If the source is modeled as $Sin(\omega t)$, the resultant generated field is:

$$Field \propto Mag \cdot Sin[\{kr + ph\} - \omega t] = Mag \cdot Sin(\theta - \omega t)$$

where: $\text{Mag} = \sqrt{\text{x}^2 + y^2}$

It should be noted that the formula describing the phase is dependent on the quadrant of the complex vector.

**Fig. 7**

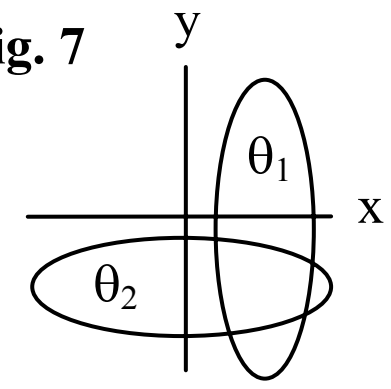

$$\theta_1 = kr + Tan^{-1}\left(\frac{y}{x}\right) \qquad \theta_2 = kr - Cos^{-1}\left(\frac{x}{\sqrt{x^2 + y^2}}\right) \tag{54}$$

**Phase speed analysis**

Phase speed can be defined as the speed at which a wave composed of one frequency propagates. The phase speed ($c_{ph}$) of an oscillating field of the form $Sin(\omega t - kr)$, in which $k = k(\omega, r)$, can be determined by setting the phase part of the field to zero, differentiating the resultant equation, and solving for $\partial r / \partial t$.

$$\frac{\partial}{\partial t}(\omega t - kr) = 0 \qquad \therefore \omega - k\frac{\partial r}{\partial t} - r\frac{\partial k}{\partial r}\frac{\partial r}{\partial t} = 0$$

$$\therefore c_{ph} = \frac{\partial r}{\partial t} = \frac{\omega}{k + r\frac{\partial k}{\partial r}} \tag{55}$$

Differentiating $\theta \equiv -kr$ with respect to r yields:

$$\frac{\partial \theta}{\partial r} = -k - r\frac{\partial k}{\partial r} \tag{56}$$

Combining these results and inserting the far-field wave number (k = ω/c) yields:

$$\boxed{c_{ph} = -\omega \bigg/ \frac{\partial \theta}{\partial r} = -c_o k \bigg/ \frac{\partial \theta}{\partial r}} \tag{57}$$

## Group speed analysis

The group speed of an oscillating field of the form: $Sin(\omega t - kr)$, in which $k = k(\omega, r)$, can be calculated by considering two Fourier components of a wave group [13]:

$$Sin(\omega_1 t - k_1 r) + Sin(\omega_2 t - k_2 r) \;=\; Sin(\Delta\omega t - \Delta k r)\; Sin(\omega t - kr) \tag{58}$$

in which: $\Delta\omega = \dfrac{\omega_1 - \omega_2}{2}$, $\Delta k = \dfrac{k_1 - k_2}{2}$, $\omega = \dfrac{\omega_1 + \omega_2}{2}$, $k = \dfrac{k_1 + k_2}{2}$

The group speed ($c_g$) can then be determined by setting the phase part of the modulation component of the field to zero, differentiating the resultant equation, and solving for $\partial r / \partial t$:

$$\frac{\partial}{\partial t}(\Delta\omega t - \Delta k r) = 0 \qquad\qquad \therefore \Delta\omega - \Delta k \frac{\partial r}{\partial t} - r \frac{\partial \Delta k}{\partial r}\frac{\partial r}{\partial t} = 0$$

$$\therefore c_g = \frac{\partial r}{\partial t} = \frac{\Delta\omega}{\Delta k + r \dfrac{\partial \Delta k}{\partial r}} \tag{59}$$

Differentiating $\Delta\theta \equiv -\Delta k r$ with respect to r yields:

$$\frac{\partial \Delta\theta}{\partial r} = -\Delta k - r \frac{\partial \Delta k}{\partial r} \tag{60}$$

Combining these results and using the far-field wave number (k = ω/c) yields:

$$c_g = -\Delta\omega \Big/ \frac{\partial \Delta\theta}{\partial r} = -\left[\frac{\partial}{\partial r}\frac{\Delta\theta}{\Delta\omega}\right]^{-1} \tag{61}$$

$$\boxed{\therefore c_g \underset{\frac{\Delta\theta}{\Delta\omega} small}{\underset{\lim}{=}} -\left[\frac{\partial^2 \theta}{\partial r \partial \omega}\right]^{-1} = -c\left[\frac{\partial^2 \theta}{\partial r \partial k}\right]^{-1}} \tag{62}$$

It should be noted that other derivations of the above phase and group speed relations are available in previous publications by the author [2, 3, 4, 5, 6, 7, 8, 9] *and in the following well-known reference* [14]. In addition, in order for the group speed to be valid, a signal should not distort as it propagates. It is known from electronic signal theory that in order to minimize signal distortion, the phase vs. frequency curve must be approximately linear over the bandwidth of the signal and the amplitude vs. frequency curve must be approximately constant over the bandwidth of the signal [15]. It is shown below that the amplitude vs. frequency curve can even be approximately linear over the bandwidth of the signal, provided the ratio of the slope of the curve to the signal amplitude is small [16 (p. 12)]. Assuming that the amplitude vs. frequency curve is increasing and approximately linear over the bandwidth of a modulated carrier signal, each signal magnitude (A) Fourier component ($w_m$) will be increased by (u) and the Fourier component symmetric about the carrier ($w_c$) will be reduced by (u):

**Fig. 8**

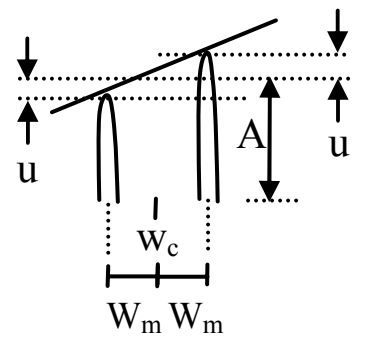

$$\frac{1}{2}\left[(A-u)\ Sin(w_c t - w_m t) + (A+u)\ Sin(w_c t + w_m t)\right]$$
$$= A \cdot Cos(w_m t) Sin(w_c t) + u \cdot Sin(w_m t) Cos(w_c t) \qquad (63)$$

The two Fourier components form an amplitude modulated signal where the magnitude of the carrier is:

$$\sqrt{A^2 \cdot Cos^2(w_m t) + u^2 \cdot Sin^2(w_m t)} \approx A \cdot Cos(w_m t) \qquad (64)$$

It should be noted that distortions to the magnitude are minimal provided: $u^2/A^2 << 1$

where (u) can be approximated using the derivative relation:

$$u = \frac{\Delta k}{2} \frac{\partial A}{\partial k}, \qquad (65)$$

provided the amplitude vs. frequency curve is approximately linear over the bandwidth of the signal. In addition, it should be noted that phase speed and group speed can also be determined from two different frequency components ($\omega_1$, $\omega_1$) using relations (Eq. 57, 62):

$$c_{ph} = -\omega_c \Big/ \frac{\partial \theta}{\partial r} \qquad c_g = -\Delta\omega \Big/ \frac{\partial \Delta\theta}{\partial r}$$

given: $\Delta\omega = \dfrac{\omega_2 - \omega_1}{2} = \omega_m$, $\quad \omega_c = \dfrac{\omega_1 + \omega_2}{2}$ , $\quad \theta = \omega t$

yields:

$$\boxed{c_{ph} = \frac{-\omega_c}{\dfrac{\partial}{\partial r}\left(\dfrac{\theta_2 + \theta_1}{2}\right)}} \qquad \boxed{c_g = \frac{-\omega_m}{\dfrac{\partial}{\partial r}\left(\dfrac{\theta_2 - \theta_1}{2}\right)}} \qquad (66)$$

Given two different frequencies, plots of the phase speed and group speed can then be determined for each field component by inserting the corresponding phase relation (Eq. 67, 70, 73). It should be noted that these relations yield the same results as (Eq. 57, 62) provided the phase and magnitude vs. frequency curves are approximately linear over the bandwidth of the signal.

## Wave propagation analysis of near-field electric dipole EM fields

To determine how the EM fields propagate in an electric dipole system, one can apply the above phase and group speed relations (Eq. 57, 62) to the known theoretical solution of an electric dipole (Eq. 47, 52, 53).

### Radial electric field ($E_r$) solution

$$y = -kr \qquad x = 1$$

$$\theta = kr - Tan^{-1}(kr) \underset{kr<<1}{\approx} -\frac{1}{3}(kr)^3 \tag{67}$$

$$c_{ph} = c_o\left(1 + \frac{1}{(kr)^2}\right) \underset{kr<<1}{\approx} \frac{c_o}{(kr)^2} \underset{kr>>1}{\approx} c_o \tag{68}$$

$$c_g = \frac{c_o\left(1+(kr)^2\right)^2}{3(kr)^2+(kr)^4} \underset{kr<<1}{\approx} \frac{c_{ph}}{3} \underset{kr>>1}{\approx} c_o \tag{69}$$

$E_r$ Phase vs. kr

**Fig. 9**

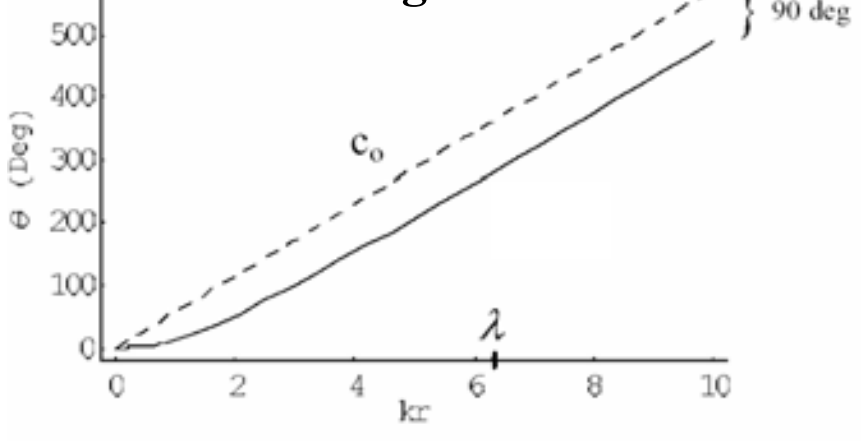

$E_r$ $c_{ph}/c_o$ vs. kr

**Fig. 10**

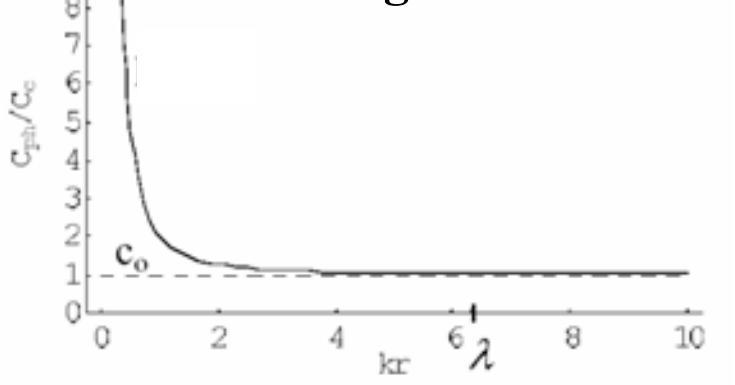

$E_r$ $c_g$ vs. kr

**Fig. 11**

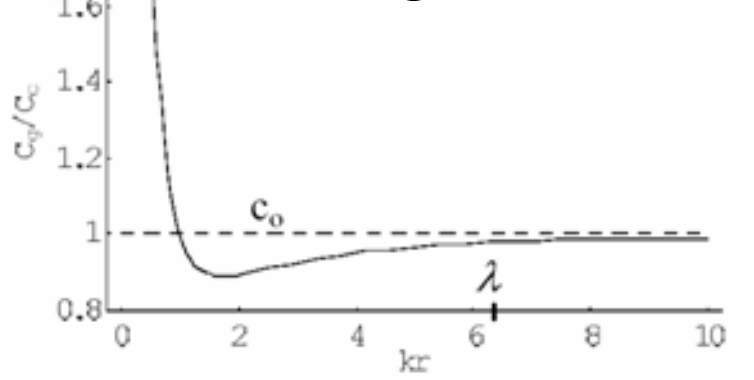

### Transverse electric field ($E_\theta$) solution

$$y = -kr \qquad x = 1-(kr)^2$$

$$\theta = kr - Cos^{-1}\left(\frac{1-(kr)^2}{\sqrt{1-(kr)^2+(kr)^4}}\right) \tag{70}$$

$$c_{ph} = c_o\left(\frac{1-(kr)^2+(kr)^4}{-2(kr)^2+(kr)^4}\right) \tag{71}$$

$$c_g = \frac{c_o\left(1-(kr)^2+(kr)^4\right)^2}{-6(kr)^2+7(kr)^4-(kr)^6+(kr)^8} \tag{72}$$

$E_\theta$ Phase vs. kr

**Fig. 12**

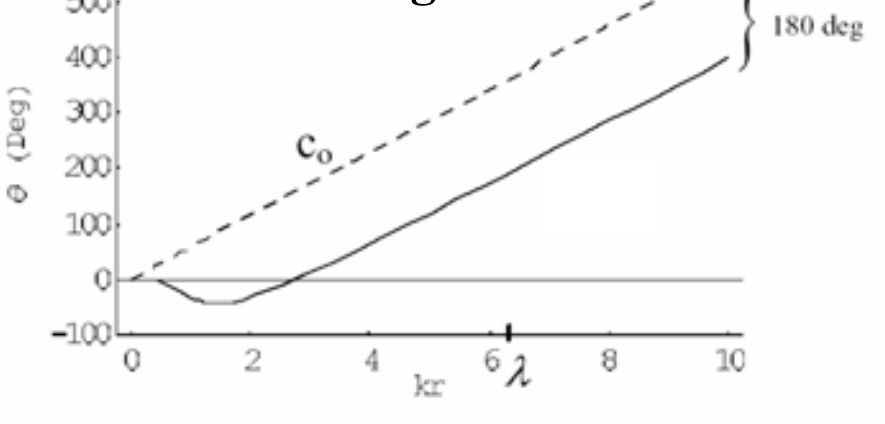

$E_\theta$ $c_{ph}/c_o$ vs. kr

**Fig. 13**

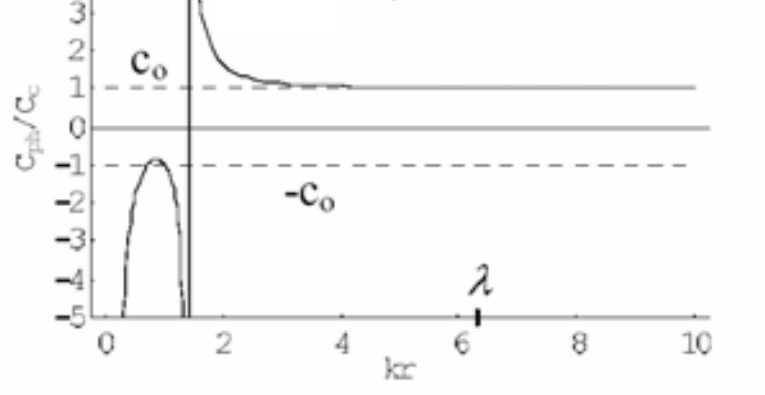

$E_\theta$ $c_g/c_o$ vs. kr

**Fig. 14**

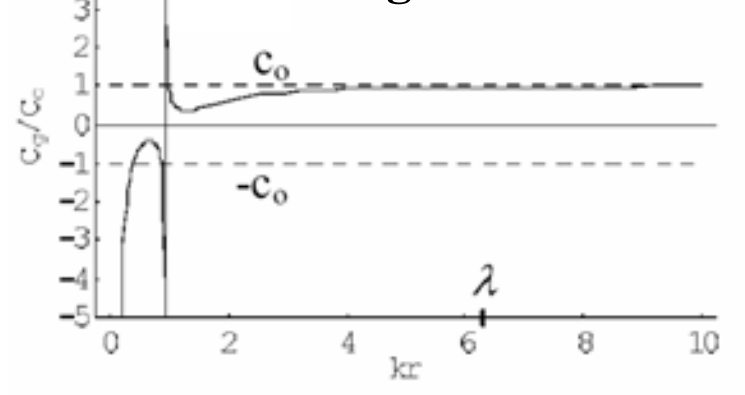

Transverse magnetic field ($B_\phi$) solution

$$y = -1 \qquad x = -kr$$

$$\theta = kr - Cos^{-1}\left(\frac{-kr}{\sqrt{1+(kr)^2}}\right) \tag{73}$$

$$c_{ph} = c_o\left(1 + \frac{1}{(kr)^2}\right) \tag{74}$$

$$c_g = \frac{c_o\left(1+(kr)^2\right)^2}{3(kr)^2 + (kr)^4} \tag{75}$$

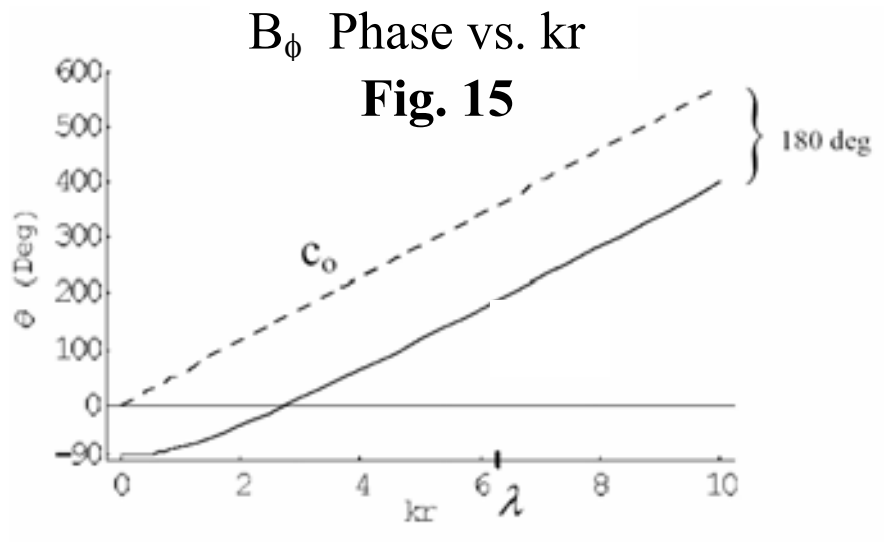

$B_\phi$ Phase vs. kr
**Fig. 15**

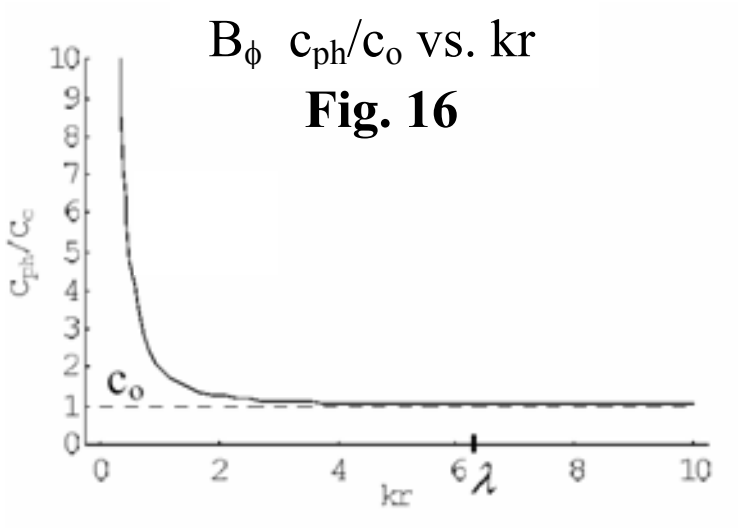

$B_\phi$ $c_{ph}/c_o$ vs. kr
**Fig. 16**

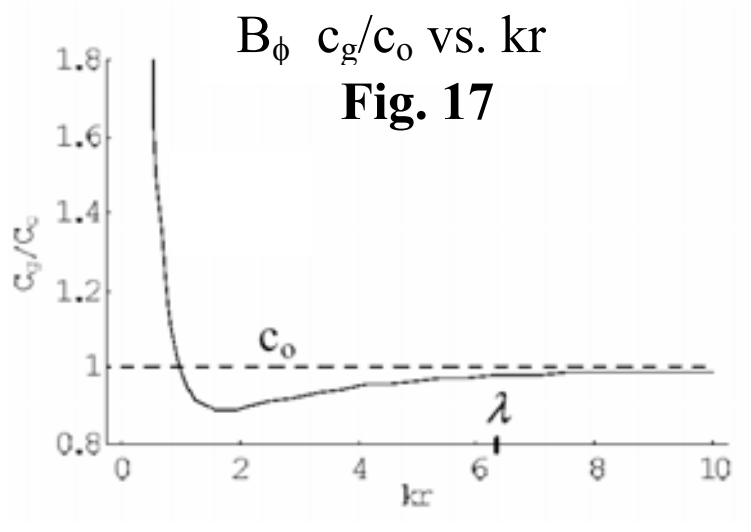

$B_\phi$ $c_g/c_o$ vs. kr
**Fig. 17**

The above results (p. 13, 14) were originally published by the author in 1999 [5], but the propagation of the longitudinal electric field and gravitational field next to a source was published earlier by the author in 1997 [6, 7]. It should be noted that after these dates similar results have been published by other authors [17, 18].

## Animated field plots

In this section, animated contour plots are presented which show how the longitudinal and transverse electric fields propagate. A cosinusoidal dipole source is used and the resultant fields are assumed to be a vectoral sum of all the wave components. The resultant field magnitude and phase are then inserted into a propagating cosine wave: Mag Cos(ωt + ph) and plotted at different moments in time. The plots are generated using Mathematica Ver. 3 software. The code generates 24 plots evenly spaced within a specified analysis period. Several of the resultant frames are shown below. The vertical dipole source is located in the center of the plots. The frames shown below (Fig. 18) are animated plots of the longitudinal electric field. They clearly show that the waves are generated at the source and propagate away from the source.

**Fig. 18** $E_r$ near-field wave animation plots

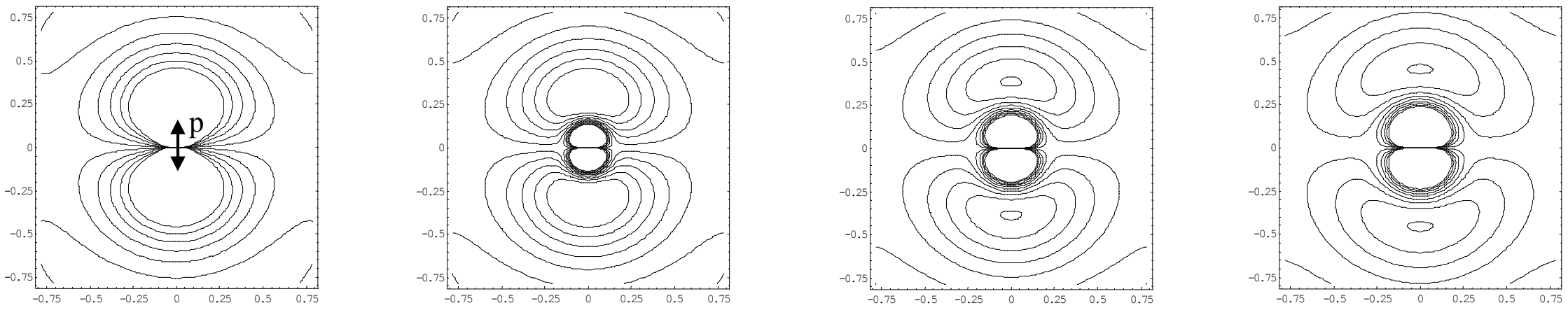

$t = \frac{17}{23}(T)$ $t = \frac{18}{23}(T)$ $t = \frac{19}{23}(T)$ $t = \frac{20}{23}(T)$

**Fig. 19** Mathematica code used to generate animations

```
<<Graphics`Animation`
Eth=MagEth*Cos[w*t+PhEth];
MagEth=po/4/Pi/eo*Sqrt[(1-(k*r)^2)^2+(k*r)^2]/r^3*Sin[th];
PhEth=-k*r+ArcCos[(1-(k*r)^2)/Sqrt[1-(k*r)^2+(k*r)^4]];
Er=MagEr*Cos[w*t+PhEr];
MagEr=po/2/Pi/eo*Sqrt[1+(k*r)^2]/r^3*Cos[th];
PhEr=-k*r+ArcTan[k*r];
L=1;k:=2*3.14159/L;c=3*10^8;w=2*3.141159*c/L;
T=L/c;po=1.6*10^(-19);eo=8.85*10^(-12);
r=Sqrt[x^2+y^2];
th=ArcCos[y/(Sqrt[x^2+y^2])];
Animate[ContourPlot[Er/(1*10^(-7)),{x,-Pi/4,Pi/4},{y,-Pi/4,Pi/4},
        PlotPoints->100],{t,0,1*T},ContourShading->False,
       Contours->{-.9,-.7,-.5,-.3,-.1,.1,.3,.5,.7,.9}]
```

The frames shown below (Fig. 20) are animated plots of the transverse electrical field (ref. Fig. 12). The plots clearly show that the waves are created outside the source and propagate toward and away from the source.

**Fig. 20** $E_\theta$ near-field wave animation plots

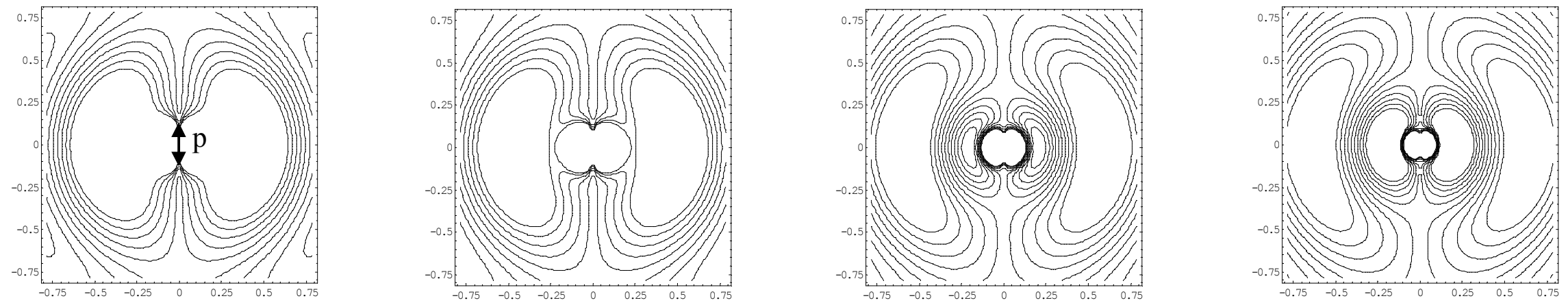

$t = \frac{1}{23}(T)$ $t = \frac{2}{23}(T)$ $t = \frac{4}{23}(T)$ $t = \frac{5}{23}(T)$

The frames shown below (Fig. 21) are animated plots of the longitudinal and transverse electrical fields vectorially added together (vector plot). The vertical dipole source is located in the middle of the left-hand side of the plot.

**Fig. 21** Animated plot of $E_r$ and $E_\theta$ vectorially added together

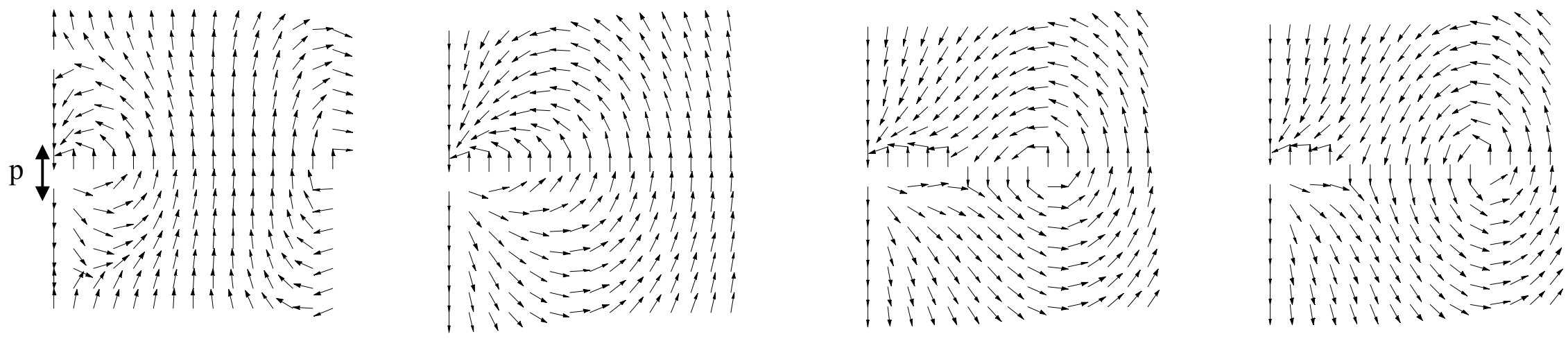

$t = \frac{7}{23}T$ $t = \frac{11}{23}T$ $t = \frac{15}{23}T$ $t = \frac{16}{23}T$

**Fig. 22** Additional Mathematica code used to generate vector plot animations
(Add this code to the previous code used above)

```
Ex=Er*Sin[th]+Eth*Cos[th];
Ey=Er*Cos[th]-Eth*Sin[th];
<<Calculus`VectorAnalysis`
<<Graphics`PlotField`
Ett={Ex,Ey};
Etmag=Sqrt[Ex^2+Ey^2];
Animate[PlotVectorField[Ett/Etmag,{x,0,0.5},{y,-0.25,0.25}],{t,0,T}]
```

A more detailed plot of the total electric field can be obtained by using the fact that a line element crossed with an electric field is zero. The resulting partial differential equation can be solved yielding [12, 20]:

Resultant equation: $$\frac{\partial}{\partial r}\left(rC_{\phi}Sin(\theta)\right)dr + \frac{\partial}{\partial \theta}\left(rC_{\phi}Sin(\theta)\right)d\theta = 0 \qquad (76)$$

Solution: $$\sqrt{1+\frac{1}{(kr)^2}}Sin^2(\theta)Cos\left[kr - Tan^{-1}(kr) - \omega t\right] = Const \qquad (77)$$

A contour plot of this solution yields the plot below (Fig. 23, compare Fig. 21).

**Fig. 23** Animated plot of E field in near-field

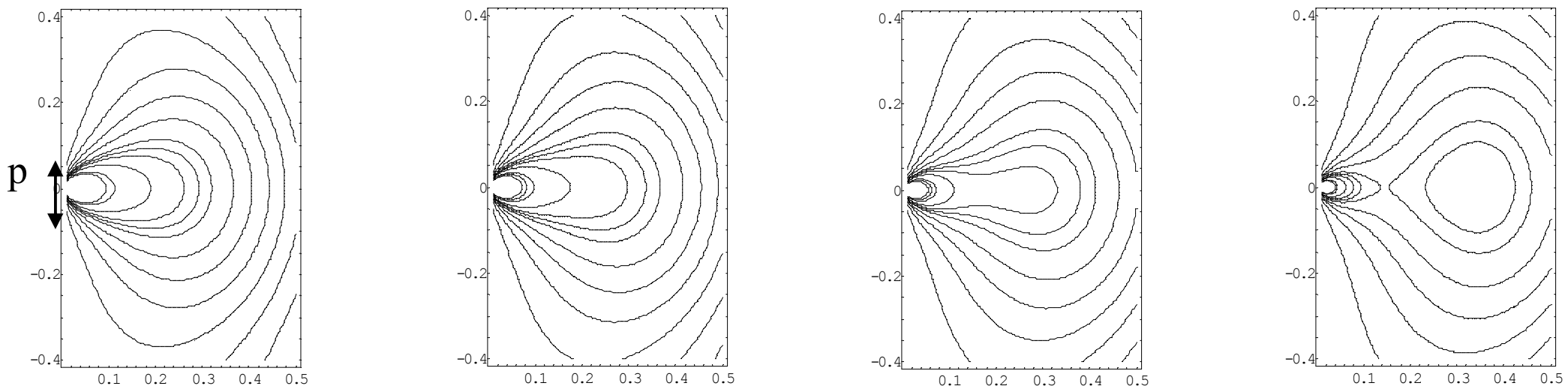

$t = \frac{1}{23}T$ $t = \frac{2}{23}T$ $t = \frac{3}{23}T$ $t = \frac{4}{23}T$

**Fig. 24** Mathematica code used to generate E field contour plot

```
L= 1;c=3*10^8; f= c/L;T=1/f;w=2*N[Pi]*f;k= w/c;
fn= Sqrt[1/ (k*r)^2+ 1] *Cos [Th]^2* Cos[w*t-k*r+ArcTan[k*r]] ;
r = Sqrt[x^2 + y^2] ;Th = ArcTan[y/x] ;
Animate[ContourPlot[fn, {x, 0.01,.5}, (y, -.4,.4), PlotPoints-> 100] ,
    {t, 0, T} , ContourShading->False,
    Contours->{-. 9, -.7, -.5, -.3, -.1, .1, .3, .5, .7, .9},
   AspectRatio->3/2]
```

Further away from the source the plot of the electric field becomes:

**Fig. 25** Animated plot of E field in farfield

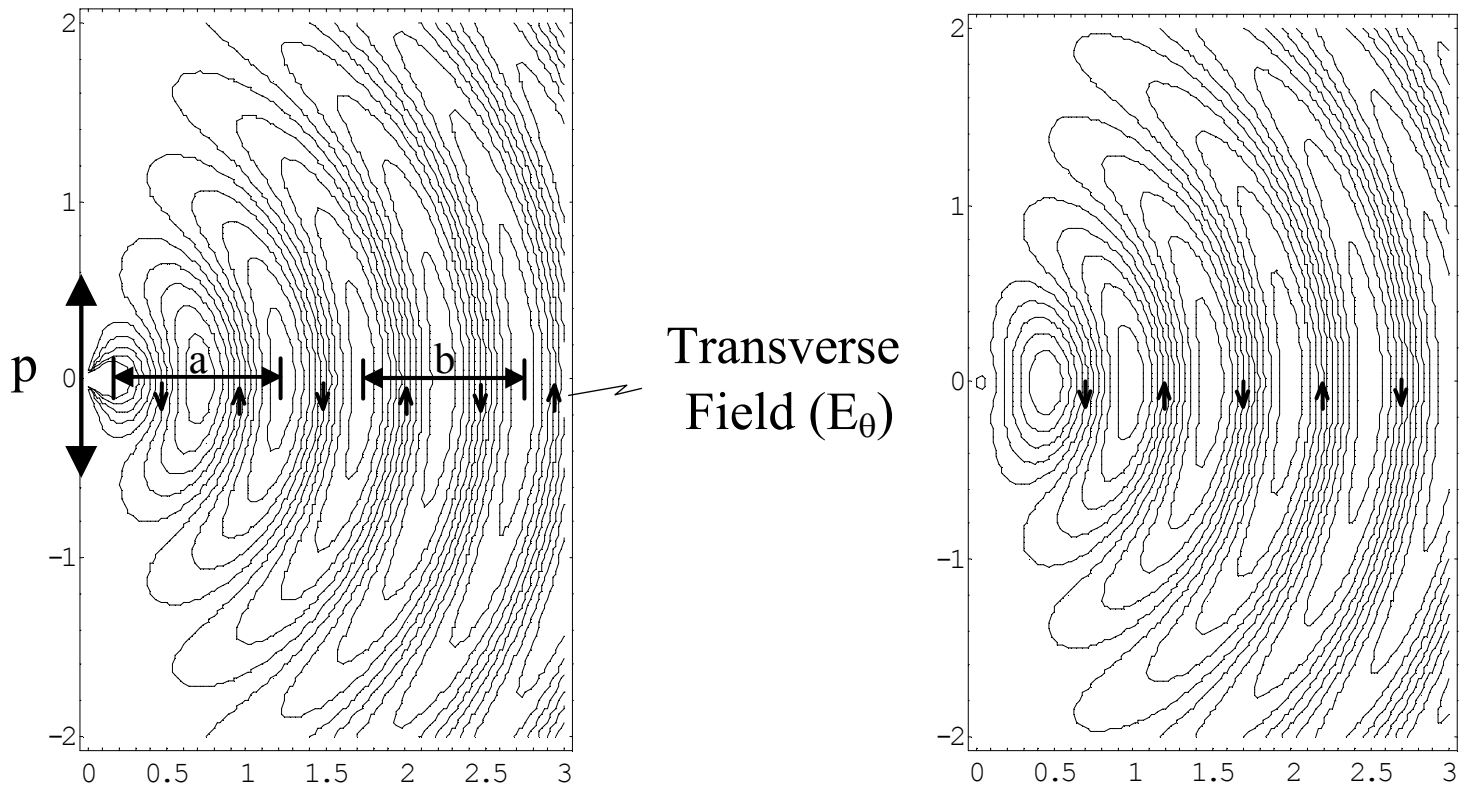

$t = \frac{0}{23}(3\,T)$ $t = \frac{3}{23}(3\,T)$

Note that careful inspection of the plot reveals that the wavelength of the transverse electric field in the near-field (a) is larger than the wavelength in the far-field (b). The phase speed (cph) is known to be a function of wavelength (λ) and frequency (f): cph = λf. Solving the relation for (f), which is constant both in the near-field and farfield, yields: f = Cphnear / λnear = Cphfar / λfar. Solving this for Cphnear yields:
Cphnear = Cphfar (λnear / λfar). Since λnear > λ far the phase speed of the transverse electric field in the nearfield is larger than the speed of light (cph > c).

## Graphical demonstration of superluminal phase and group speed

### Superluminal near-field phase speed of radial electric field

To demonstrate the superluminal near-field phase velocity of the longitudinal electric field, the calculated phase and amplitude functions can be inserted into a cosine signal and the field amplitude can then be plotted in the near field as a function of space (r) at several isolated moments in time (t), (Fig. 27). A field propagating at the speed of light (shown as a dashed line) is also included in the plot for reference. The following parameters are used in the subsequent plots: 1m wavelength ($\lambda$), 300GHz signal frequency (f), 3.3ns signal period (T). The following Mathematica code (Fig. 26) is used to generate these plots:

**Fig. 26** Mathematica code used to generate plots

```
L=1;c=3*10^8;f=c/L;T=1/f;w=2*N[Pi]*f;k=w/c;
Animate[Plot[{1/r^2*Sqrt[1/r^2*k^2]
        *Cos[w*t-k*r+ArcTan[k*r]],20*Cos[w*t-k*r]},
         {r,0.1,3*L},PlotPoints->600,
         PlotRange->{-60,60}],{t,0,3*T}];
```

**Fig. 27** $E_r$ vs Space – Cosinusoidal Signal

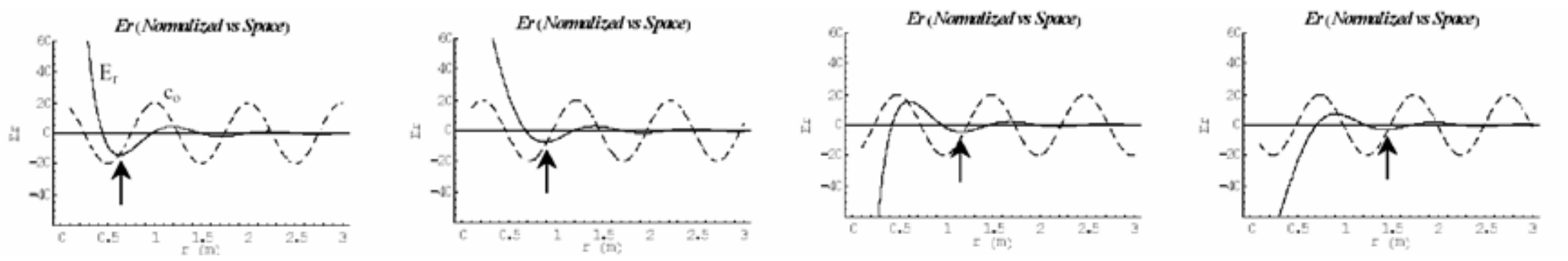

$t = \frac{0}{23}(T)$ $t = \frac{5}{23}(T)$ $t = \frac{11}{23}(T)$ $t = \frac{17}{23}(T)$

The longitudinal field (shown as a solid line in the plot above) is observed to propagate away from the source, which is located at r = 0. As it propagates away from the source, the oscillation amplitude decays rapidly ($1/r^3$) near the source ($r < \lambda$), and decreases more slowly ($1/r^2$) in the farfield ($r > \lambda$) (ref. Eq. 47, 52, 53). A field propagating at the speed of light (shown as a dashed line in the plot above) is also included in the plot for reference. Both signals start together in phase. The longitudinal field is seen to propagate faster than the light signal initially when it is generated at the source. After propagating about one wavelength the longitudinal electric field is observed to slow down to the speed of light, resulting in a final relative phase difference of 90 degrees. In order to see the effect more clearly the signals can be plotted with the amplitude part of the function set to unity (Fig. 28).

**Fig. 28** $E_r$ (Normalised) vs Space – Cosinusoidal Signal

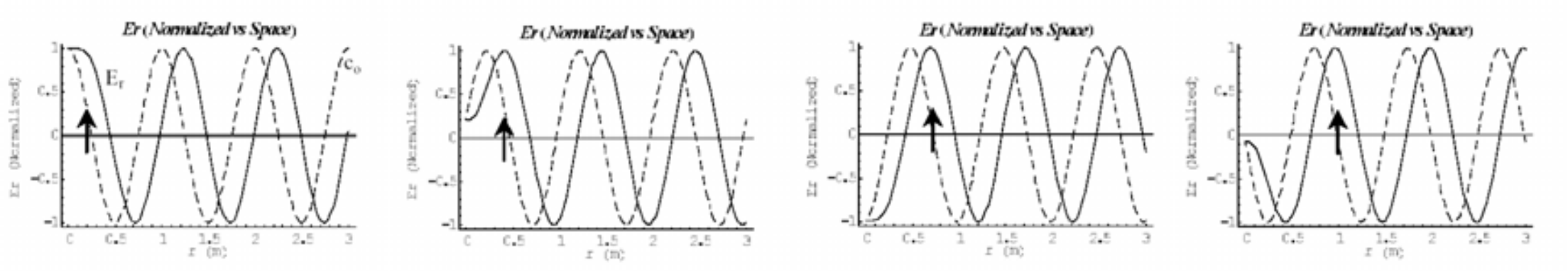

$t = \frac{0}{23}(T)$ $t = \frac{5}{23}(T)$ $t = \frac{11}{23}(T)$ $t = \frac{17}{23}(T)$

It is also instructive to plot the signals as a function of time (t) for several positions (r) away from the source (Fig 29). At the source (r = 0) both signals are observed to be in phase. Further away from the source the longitudinal field signal is observed to shift 90 degrees, indicating that it arrives earlier in time. The plots shown below are normalized for clarity, but it should be noted that the signals have the same form even if the amplitude part of the function were included. The only difference is the vertical scaling of the plot. From these plots it can also be seen that the longitudinal field propagates much faster than the speed of light near the source ($r < \lambda$), and reduces to the speed of light at about one wavelength from the source ($r \cong \lambda$), resulting in a final relative phase difference of 90 degrees between the longitudinal field (shown as a solid line), and the field propagating at the speed of light (shown as dashed line).

**Fig. 29** Er (Normalised) vs Time – Cosinusoidal Signal

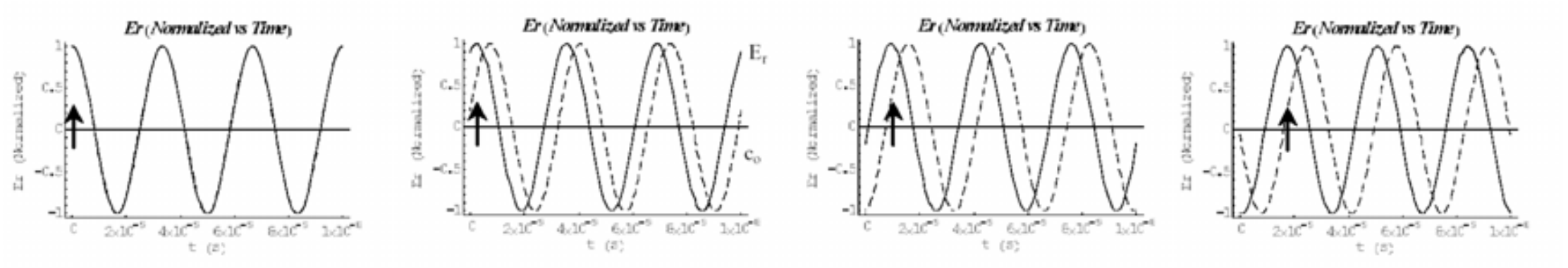

$r = \frac{0}{23}(\lambda)$ $r = \frac{5}{23}(\lambda)$ $r = \frac{11}{23}(\lambda)$ $r = \frac{17}{23}(\lambda)$

Superluminal near-field group speed of radial electric field

To demonstrate the superluminal near-field group propagation speed of the longitudinal field, the calculated phase and amplitude functions can be inserted into the spectral components of an amplitude modulated cosine signal, and the field amplitude can then be plotted as a function of space (r) at several isolated moments in time (t), (Fig. 31). To demonstrate this technique the group propagation (shown as a solid line) is compared to the phase speed propagation (shown as a dashed line) of waves of the form: Cos(kr - wt). Note that the phase component (kr) is independent of frequency. This result is known to produce group waves and phase waves that both propagate at the speed of light. This can be seen by using (Eq. 14): since $\theta$ = kr $\therefore c_{ph} = \omega / \frac{\partial \theta}{\partial r} = \omega / k = c$. Using (Eq. 62) $c_g = c\left[\partial\theta^2 / \partial r \partial k\right]^{-1} = c$. The following parameters are used in the following plots: Carrier part of signal [1m wavelength ($\lambda_c$), 300MHz signal frequency (fc), 3.3ns signal period (Tc)] , Modulation part of the signal [10m wavelength, 30MHz signal frequency (fm), 33.3ns signal period (Tm)]. Note that the phase relation of both signals are the same at different isolated moments in time and that both signals propagate away from the source at the same speed. The following Mathematica code (Fig. 30) is used to generate these plots:

**Fig. 30** Mathematica code was used to generate these plots

```
AM=Cos[Wc*t]*(1+Cos[Wm*t]);ph1=kc*x;ph2=(kc-km)*x; ph3=(kc+km)*x;
Am1=TrigReduce[AM];
AM2=1/2*(2 Cos[t*Wc-ph1]+Cos[t*Wc-t*Wm-ph2]+Cos[t*Wc+t*Wm-ph3]);
L=1;c=3*10^8;fc=c/L;fm=fc/10;T=1/fc;Wc=2 N[Pi] fc;Wm=2 N[Pi] fm;
kc=Wc/c;km=Wm/c;
Animate[Plot[{AM2,2*Cos[Wc*t-kc*x]},{x,0,10*L}],{t,0,10*T}];
```

**Fig. 31** Light Phase (Cosine Wave) and Group (AM Signal) vs Space

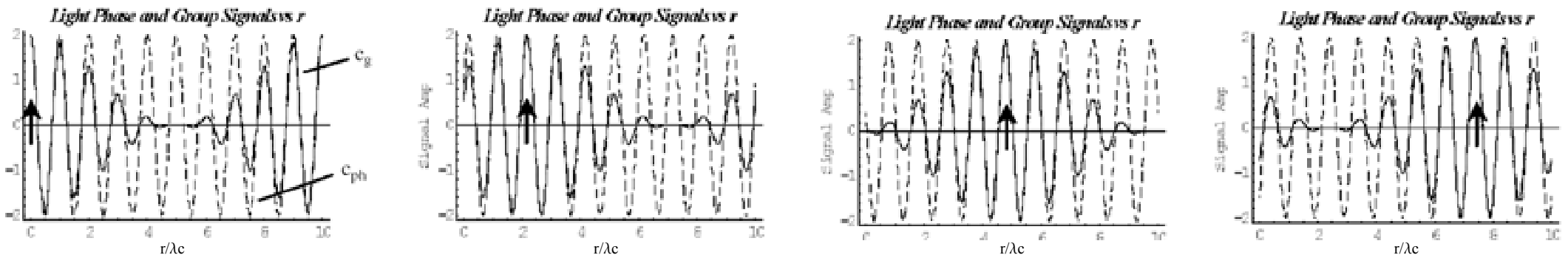

$t = \frac{0}{23}(10 \cdot T_c)$ $t = \frac{5}{23}(10 \cdot T_c)$ $t = \frac{11}{23}(10 \cdot T_c)$ $t = \frac{17}{23}(10 \cdot T_c)$

Plotting the signals as a function of time for several spatial positions from the source also shows that group and phase signals travel at the same speed and remain in phase as they propagate.

**Fig. 32** Light Phase (Cosine Wave) and Group (AM Signal) vs Time

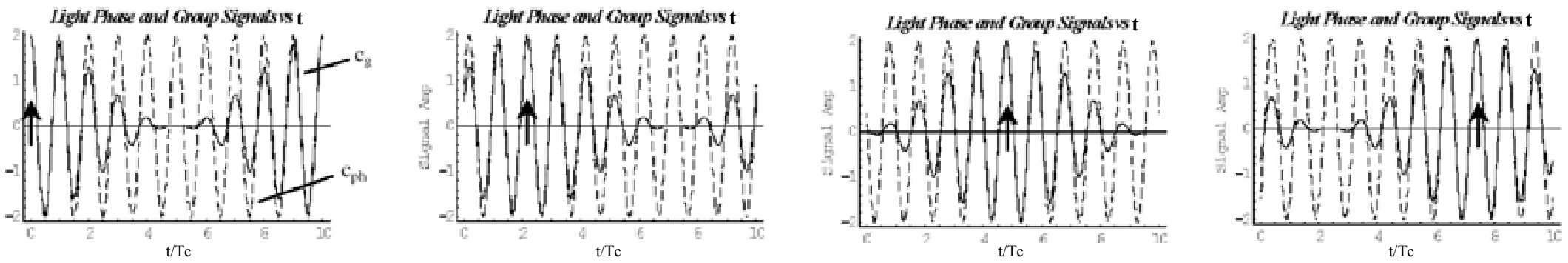

$r = \frac{0}{23}(10 \cdot \lambda)$ $\quad r = \frac{5}{23}(10 \cdot \lambda)$ $\quad r = \frac{11}{23}(10 \cdot \lambda)$ $\quad r = \frac{17}{23}(10 \cdot \lambda)$

The superluminal near-field group propagation speed of the longitudinal electric field can also be demonstrated in the same way as in the above example. The calculated phase function for the field can be inserted in into the spectral components of an amplitude modulated cosine signal and the field amplitude (shown as a solid line in the plot below) can then be plotted as a function of space (r) at several isolated moments in time (t), (Fig. 34, 35). An amplitude modulated field propagating at the speed of light (shown as a dashed line) is also included in the plot for reference (envelope propagates at speed of light). Note that for this reference signal both the phase speed and the group speed are equal to the speed of light (ref. Fig. 31, 32). The following mathematica code (Fig. 33) is used to generate the plots below. The same signal parameters used in the previous example are used in the calculation.

**Fig. 33** Mathematica code used to generate plots

```
AM= Cos[Wc t] * (1 + Cos[Wm t]) ; ph1 = kc x; ph2 = (kc - km) x; ph3 = (kc+ km) x;
ph11 = kc x+ ArcTan[-kc x] ;
ph22 = (kc- km) x + ArcTan[-(kc- km) x] ; ph33 = (kc+ km) x + ArcTan[- (kc+ km) x] ;
AM1 = TrigReduce[AM] ; AM2 = 1 / 2* (2 Cos[t Wc- ph1] + Cos[t Wc- t Wm - ph2] + Cos[t Wc+ t Wm - ph3]) ;
AM3 = 1 / 2* (2 Cos[t Wc- ph11] + Cos[t Wc- t Wm - ph22] + Cos[t Wc+ t Wm - ph33]) ;
L= 1; c= 3*10^8; fc= c/ L; fm = fc/ 10; T = 1 / fc; Wc= 2 N[Pi] fc; Wm = 2 N[Pi] fm;
kc= Wc/ c; km = Wm / c; Animate[Plot[{AM3, AM2}, {x, 0, 10* L}, PlotPoints -> 600] , {t, 0, 10* T}] ;
```

**Fig. 34** Er (Normalized) vs Space – AM Signal

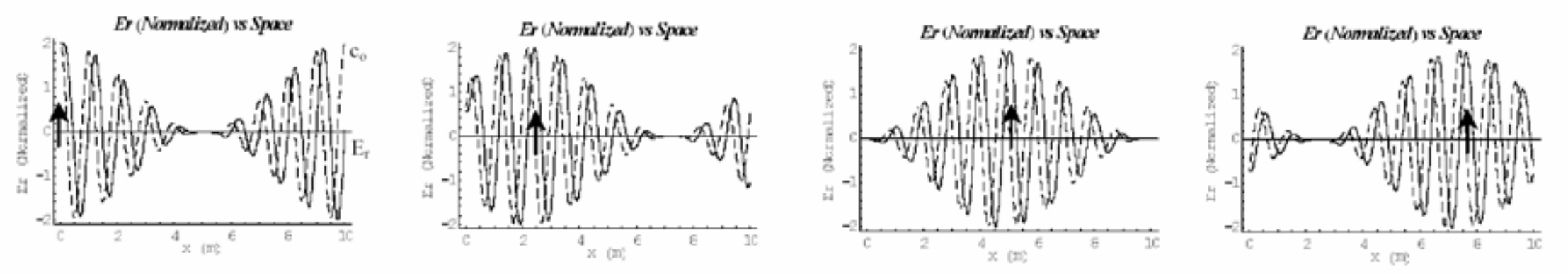

$t = \frac{0}{23}(10 \cdot T_c)$ $\quad t = \frac{5}{23}(10 \cdot T_c)$ $\quad t = \frac{11}{23}(10 \cdot T_c)$ $\quad t = \frac{17}{23}(10 \cdot T_c)$

**Fig. 35** Zoom of $E_r$ (Normalized) vs Space – AM Signal

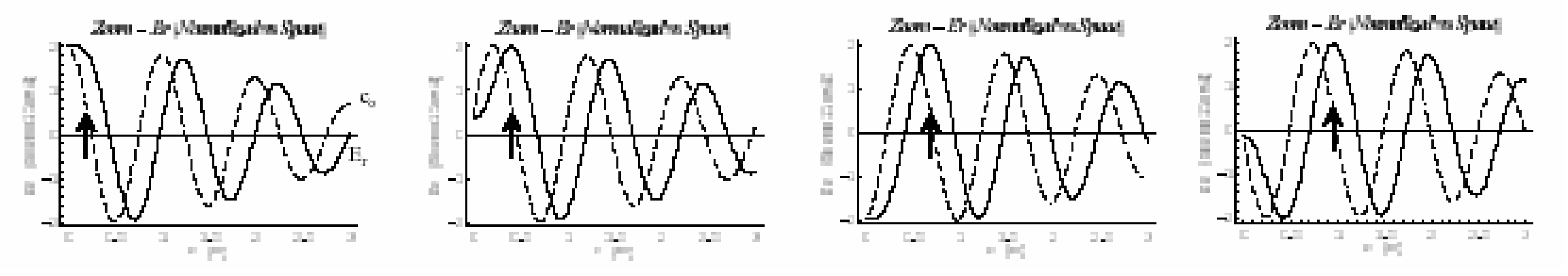

$t = \frac{0}{23}(T)$ $t = \frac{5}{23}(T)$ $t = \frac{11}{23}(T)$ $t = \frac{17}{23}(T)$

The above plots show an amplitude modulated longitudinal field group packet (shown as a solid line) propagating away from the source, at r = 0 (group maxima marked by vertical arrow). A propagating speed of light group wave (shown as a dashed line) is also provided for reference. The group maxima of the amplitude modulated longitudinal field is observed to propagate to the right side of the plot before the group maxima of the speed of light wave. These series of plots clearly demonstrate that the longitudinal group wave propagates much faster than the speed of light near the source ($r < \lambda_c$). After propagating about one carrier wavelength from the source ($r \cong \lambda_c$) the modulation part of longitudinal field (envelope of solid line) reduces to the speed of light, resulting in a final relative phase difference of 90 degrees (relative to the carrier signal) between the longitudinal field, and the field propagating at the speed of light.

It is also very instructive to plot the field amplitude of the amplitude modulated longitudinal wave (shown as a solid line in plot below) as a function of time (t), for several positions away from the source (r), (Fig. 36, 37). As before, an amplitude modulated wave traveling with light speed is also plotted for reference (shown as a dashed line). At the source (r = 0) both signals are observed to be in phase. Further away from the source the modulated longitudinal field signal is observed to shift to the left, indicating that the modulation part of longitudinal field (envelope) arrives earlier in time. The plots shown below are normalized for clarity, but it should be noted that the signals have the same form even if the amplitude part of the function were included. The only difference is the vertical scaling of the plot. From these plots it can be seen that the modulation part of longitudinal field (envelope of solid line) propagates much faster than the modulated light speed signal (envelope of dashed line propagates at speed of light) near the source ($r < \lambda_c$). After propagating about one carrier wavelength from the source ($r \cong \lambda_c$) the modulation part of longitudinal field (envelope of solid line) reduces to the speed of light, resulting in a final relative phase difference of 90 degrees (relative to the carrier signal) between the longitudinal field, and the field propagating at the speed of light.

**Fig. 36** $E_r$ (Normalized) vs Time – AM Signal

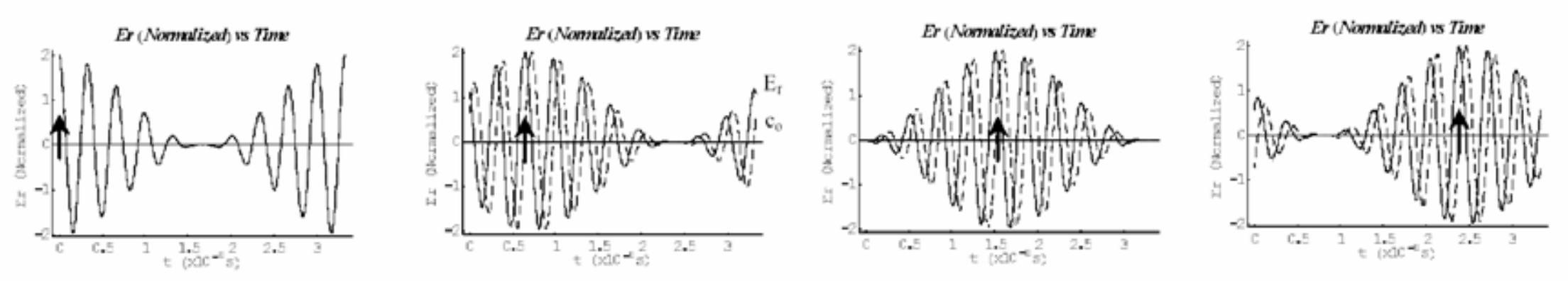

$r = \frac{0}{23}(10 \cdot \lambda)$ $r = \frac{5}{23}(10 \cdot \lambda)$ $r = \frac{11}{23}(10 \cdot \lambda)$ $r = \frac{17}{23}(10 \cdot \lambda)$

**Fig. 37** Zoom of Er (Normalized) vs Time – AM Signal

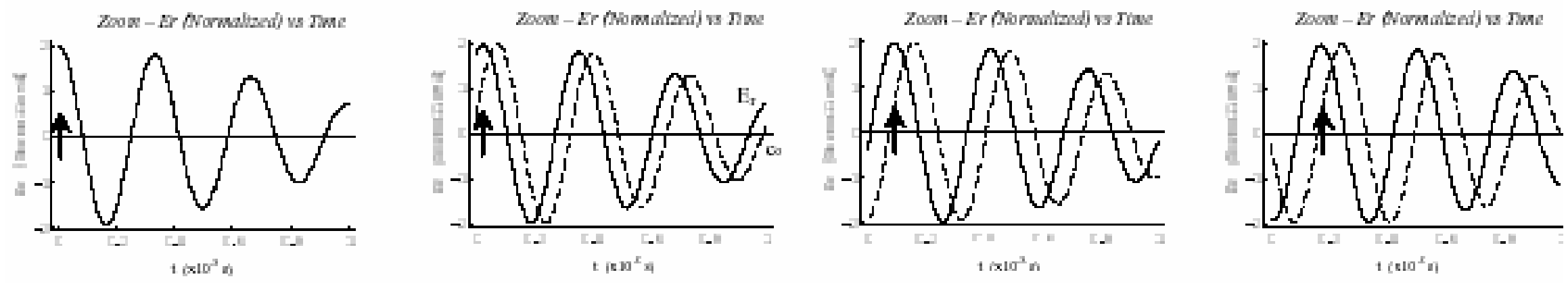

$r = \frac{0}{23}(\lambda)$ $r = \frac{5}{23}(\lambda)$ $r = \frac{11}{23}(\lambda)$ $r = \frac{17}{23}(\lambda)$

The first frame of the plot shows the two wave groups starting in phase at the source (r = 0). The following frames show that as the two waves propagate away from the source, the group maxima (marked by a vertical arrow) of the amplitude modulated longitudinal arrives earlier in time than the group maxima of a light speed amplitude modulated signal, thus demonstrating that the group speed of an amplitude modulated longitudinal electrical field is much faster than the speed of light near the source ($r < \lambda_c$), and reduces to the speed of light after it has propagated about one wavelength from the source ($r \cong \lambda_c$), resulting in a final phase difference of 90 degrees.

**<u>Interpretation of theoretical results</u>**

The above theoretical results (ref. Fig. 9 - 17) suggest that longitudinal electric field waves and transverse magnetic field waves are generated at the dipole source and propagate away. Upon creation, the waves (phase and group) travel with infinite speed and then rapidly reduce to the speed of light after they propagate about one wavelength away from the source. In addition, transverse electric field waves (phase and group) are generated approximately one-quarter wavelength outside the source and propagate toward and away from the source. Upon creation, the transverse waves travel with infinite speed. The outgoing transverse waves reduce to the speed of light after they propagate about one wavelength away from the source. The inward propagating transverse fields rapidly reduce to the speed of light and then rapidly increase to infinite speed as they travel into the source. In addition, the above results show that the transverse electrical field waves are generated about 90 degrees out of phase with respect to the longitudinal waves. In the near-field the outward propagating longitudinal waves and the inward propagating transverse waves combine together to form a type of oscillating standing wave. Note that unlike a typical standing wave the outward and inward waves are completely different types of waves (longitudinal vs. transverse) and can be separated by proper orientation of a detecting antenna. In addition, it should also be noted that both the phase and group waves are not confined to one side of the speed of light boundary and propagate at speeds above and below the speed of light in specific regions from the source.

The mechanism by which the electromagntetic near-field waves become superluminal can be understood by noting that the field components can be considered rectangular vector components of the total field (Fig. 38). For example, the vector diagram for the longitudinal electric field (ref. Eq. 56) is:

**Fig. 38** Vector diagram for longitudinal electric field

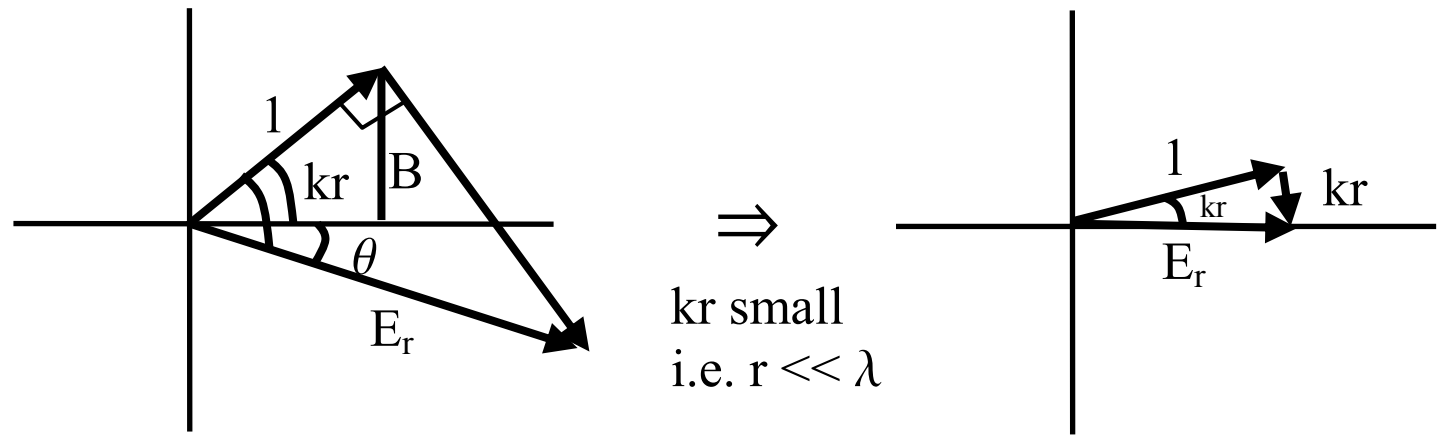

From this vector diagram it can be seen that the phase of the longitudinal electric field is: $\theta = \phi$ - kr. Also it can be seen that angle: $\phi$ = ArcTan[kr]. Combining these relations yields phase relation (Eq. 67): $\theta$ = ArcTan[kr] – kr. Note that for small (kr $\therefore$ r $\ll \lambda$) the angle bisector: B = 1 Sin(kr) $\cong$ kr has about the same length as vector (kr). Therefore when the (kr) is small the two vector components add together to form a longitudinal electric field vector which has nearly zero phase. Note that the angle bisector approximation is valid for several values of (kr) when (kr) is small. This result can also be seen by Taylor expanding the phase relation for small (kr) yielding:
$\theta = kr - [kr + (kr)^3/3 + O(kr)^5] = (kr)^3/3 + O(kr)^5$, where $kr = \omega r/c$. These results show that very near the dipole source the phase of the longitudinal electric field is zero, causing both the phase speed and the group speed to be infinite (ref. Eq. 57, 62). In the near-field the phase increases to $(kr)^3/3$, causing the phase speed to be: $c/(kr)^2$ (ref. Eq. 57) and the group speed to be: $c/(kr)^2/3$ (ref. Eq. 62). In the far-field the phase becomes: $\pi$ - kr, causing both the phase speed and the group speed to be equal to the speed of light (ref. Eq. 57, 62). The other components of the electromagnetic field ($E_\theta$, $B_\phi$) can also be analyzed in the same way yielding similar results.

## Numerical verification of field component propagation

To verify the predicted wave propagation effects, a numerical simulation was performed. The simulation consisted of extracting the transfer function from the various field components of the known dipole solution (Eq. 47, 52, 53) and then inverse Fourier transforming ($FT^{-1}$) the Fourier transform (FT) of a given signal multiplied by the dipole transfer function [19]: Result Sig = $FT^{-1}$ [ FT [Signal] x G ], where "Result Sig" is the resultant propagating signal as a function of time, and "G" is the transfer function of the wave propagation system. A simple amplitude modulated signal (300MHz carrier, 20MHz modulation), generated by adding together two sinusoidal oscillations (280MHz and 320MHz), was applied as an input signal. The simulations yielded propagating signal-versus-time animations as a function of distance (r) from the source. To compare the resulting simulations to theoretical expectations, a propagating modulation envelope of the AM signal was superimposed: Mod = $Cos[w_m t-(\theta_2-\theta_1)/2]$ where $\theta_1$ and $\theta_2$ are the theoretically expected phase shifts for the 2 frequencies used to create the AM signal (Eq. 67, 70, 73). The results below show a very good match between theory and numerical simulation.

**Fig. 39**

Resultant $E_r$ vs. time animation plots as a function of distance (r) from source

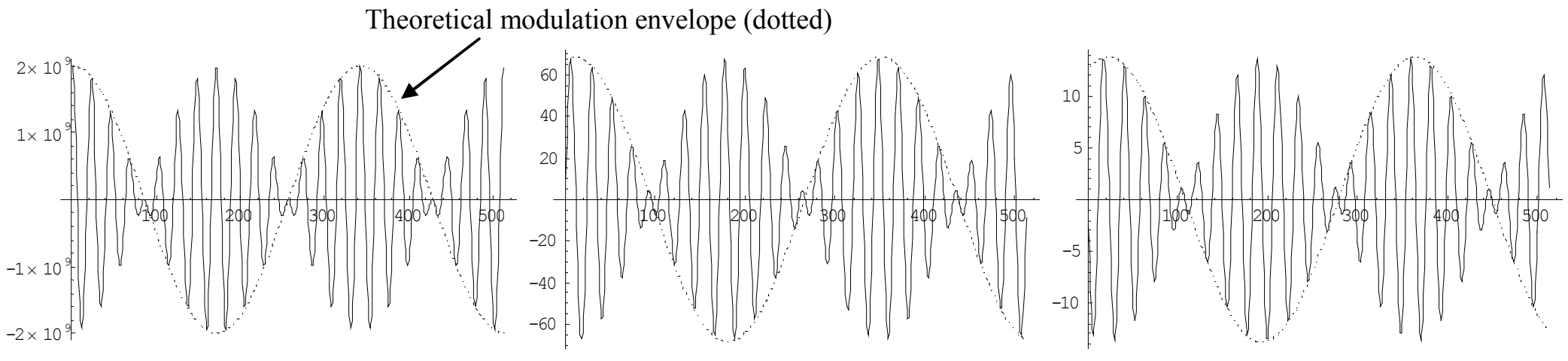

(1/25)$\lambda_c$ (12/25)$\lambda_c$ (25/25)$\lambda_c$

**Fig. 40**

Resultant $E_\theta$ vs. time animation plots as a function of distance (r) from source

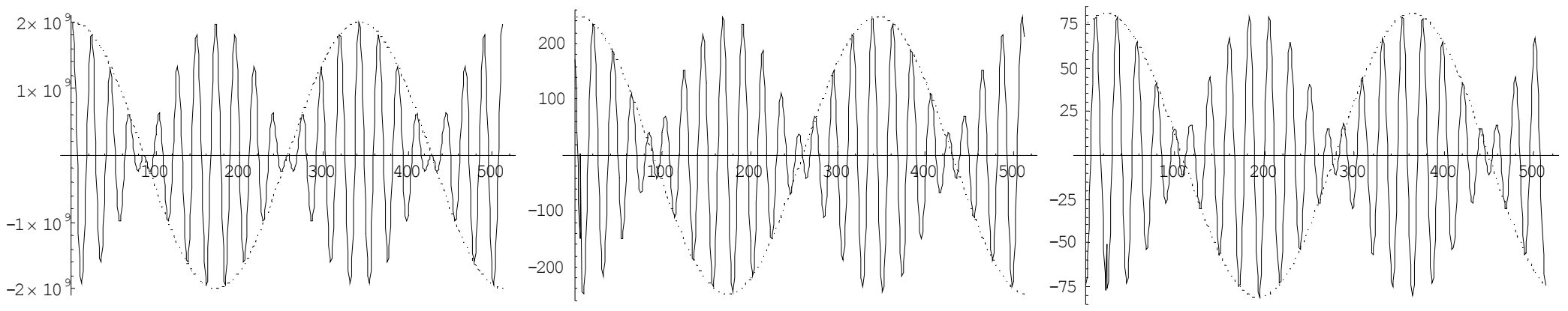

(1/25)$\lambda_c$ (8/25)$\lambda_c$ (25/25)$\lambda_c$

**Fig. 41**

Resultant $H_\phi$ vs. time animation plots as a function of distance (r) from source

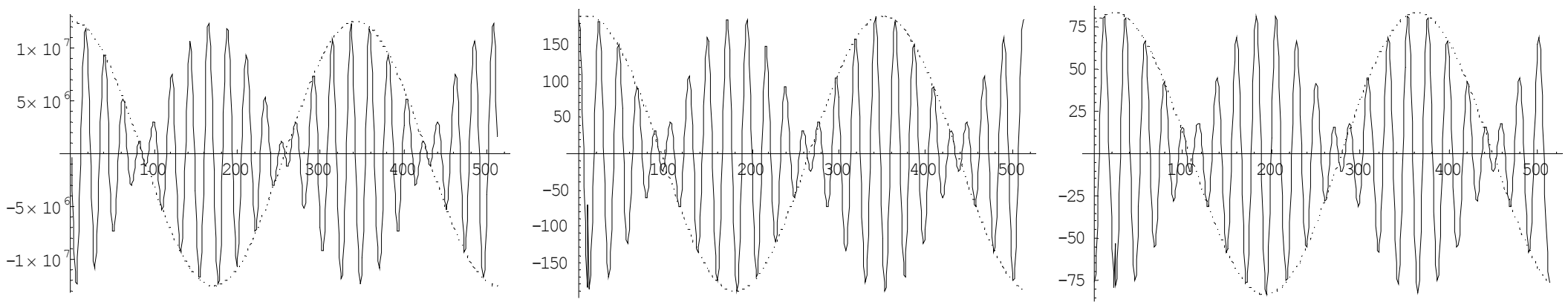

(1/25)$\lambda_c$ (12/25)$\lambda_c$ (25/25)$\lambda_c$

**Fig. 42**

Mathematica code for simulation of $E_\theta$ wave propagation

```
f1=280*10^6; f2=320*10^6; fc=(f1+f2)/2; fm=(f2-f1)/2; w1=2*Pi*f1; w2=2*Pi*f2; T=1/fm;
x=Cos[w1*t]+Cos[w2*t];        (* Input signal *)
n=1024; Cycle=1.5; ts=Cycle*T/n; fs=1/ts; fn=fs/2; v=N[Table[x,{t,0,Cycle*T,Cycle*T/n}]];
y=Fourier[v]*1.4; k=2*Pi/L; L=c/f; c=3*10^8; u=N[Table[k,{f,0,fs,fs/n}]];
h=y*(1-(u*r)*(u*r)-I*(u*r))/r^3*Exp[I*u*r];       (* FT [Signal] x G *)
l=Take[h,n/2]; <<Graphics`Animation` <<Graphics`MultipleListPlot` L1=c/f1; L2=c/f2; k1=2*Pi/L1;
k2=2*Pi/L2; km=(k2-k1)/2; kc=(k2+k1)/2; wm=2*Pi*fm; wc=2*Pi*fc;
ph1=k1*r-ArcCos[(1-(-k1*r)^2)/Sqrt[(1-(k1*r)^2)+(k1*r)^4]];      (* Eθ phase relation for f1 *)
ph2=k2*r-ArcCos[(1-(-k2*r)^2)/Sqrt[(1-(k2*r)^2)+(k2*r)^4]];      (* Eθ phase relation for f2 *)
tn=N[Table[t,{t,0,Cycle*T,Cycle*T*2/n}]];
Md=2*Abs[(1-(wc/c*r)*(wc/c*r)-I*(wc/c*r))/r^3]*Cos[wm*tn-(ph2-ph1)/2];  (* Theoretical cg plot *)
Animate[MultipleListPlot[Re[InverseFourier[l]],Md,PlotJoined->True,
   SymbolShape\[Rule]None],{r,0.001,c/fc,c/fc/25}]
```

Er animations were generated using the above code and substituting the following known Er relations

```
h=y*(1-I*(u*r))/r^3*Exp[I*u*r];    (* FT [Signal] x G *)
ph1=k1*r-ArcTan[k1*r];   (* Er phase relation for f1 *)
ph2=k2*r-ArcTan[k2*r];   (* Er phase relation for f2 *)
Md=2*Abs[(1-I*(wc/c*r))/r^3]*Cos[wm*tn-(ph2-ph1)/2];    (* Theoretical c_g plot *)
```

Bϕ animations were generated using the above code and substituting the following known Bϕ relations

```
h=y*(-(u*r)-I)*u/r^2*Exp[I*u*r];     (* FT [Signal] x G *)
ph1=k1*r-ArcCos[-k1*r/Sqrt[1+(k1*r)^2]];    (* B_ϕ phase relation for f1 *)
ph2=k2*r-ArcCos[-k2*r/Sqrt[1+(k2*r)^2]];    (* B_ϕ phase relation for f2 *)
Md=2*Abs[(-(wc/c*r)-I)]*wc/c/r^2*Cos[wm*tn-(ph2-ph1)/2];    (* Theoretical c_g plot *)
```

To check the simulator, the AM signal was also applied to a light propagating system. The results yielded a 300 MHz carrier, 20 MHz modulated AM signal animation which linearly increased its phase shift as expected. This was verified by substituting the following known relations:

```
h=y*Exp[I*u*r];     (* FT [Signal] x G *)
Md=2*Cos[wm*tn-( k2*r - k1*r)/2];     (* Theoretical c_g plot *)
```

It should be noted that no signal distortion was observed in these simulations. This can be attributed to the fact that the phase and amplitude vs. frequency curves are approximately linear over the bandwidth of the signal ($\Delta f/f_c$ = 40/300 = 1/7.5). The use of linearity constraint can be seen to be justified by plotting $u^2/A^2$ for each field component and noting that it is much less than one over the bandwidth of the signal (Eq. 65, Fig. 43-45).

$E_r$ ($u^2/A^2$ vs. r)
**Fig. 43**

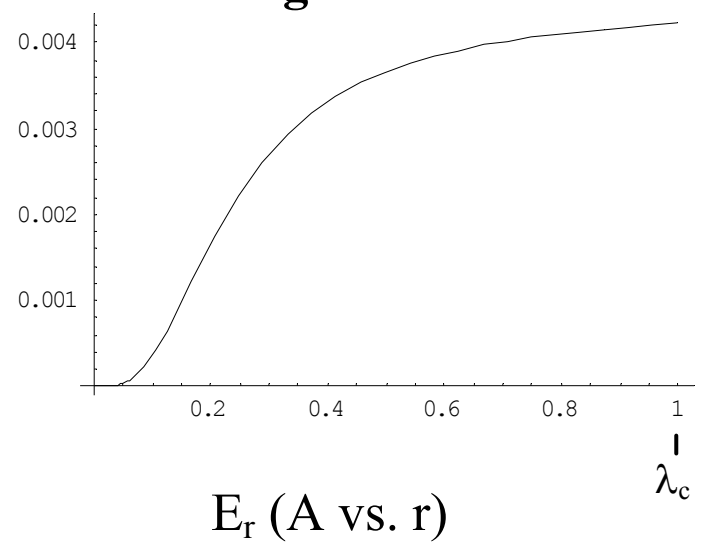

$E_\theta$ ($u^2/A^2$ vs. r)
**Fig. 44**

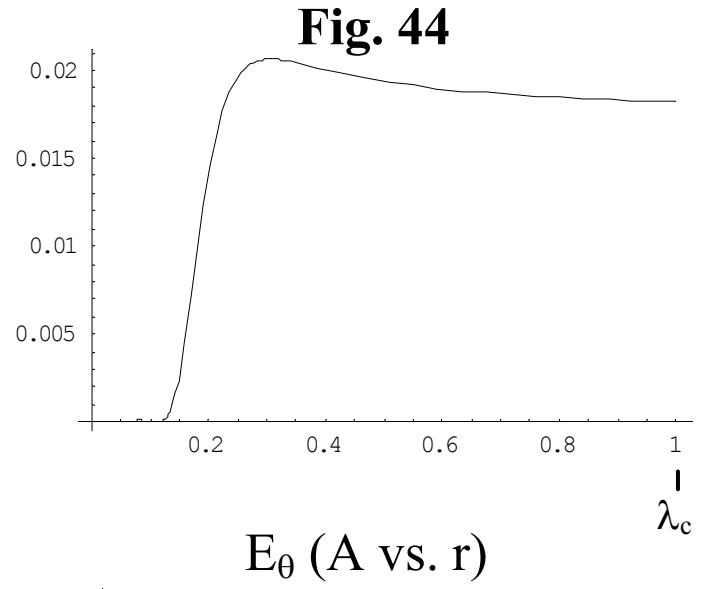

$B_\varphi$ ($u^2/A^2$ vs. r)
**Fig. 45**

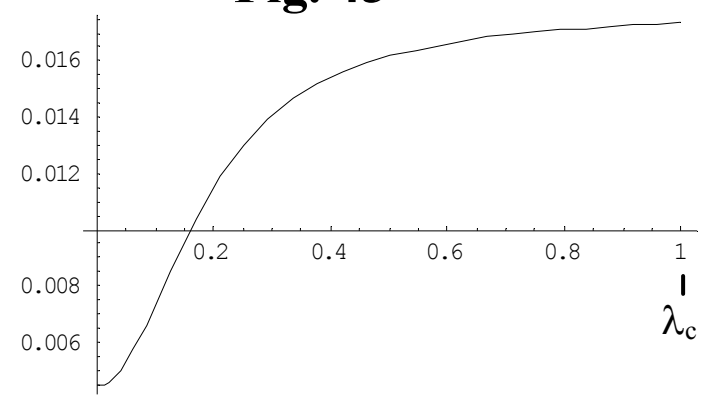

$E_r$ (A vs. r)
**Fig. 46**

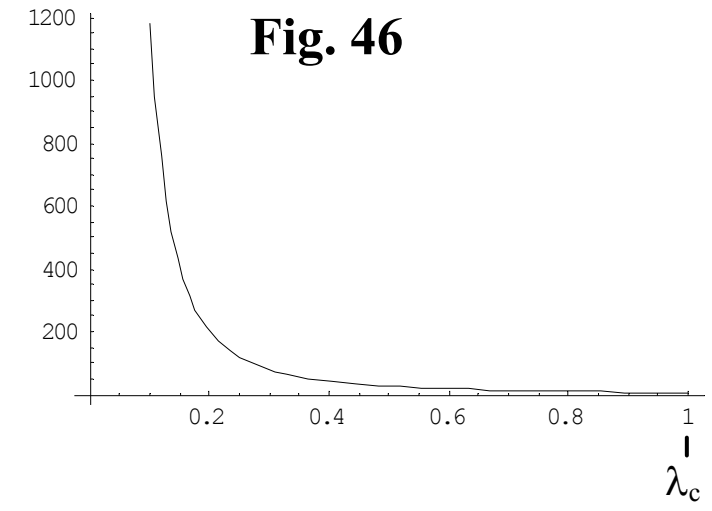

$E_\theta$ (A vs. r)
**Fig. 47**

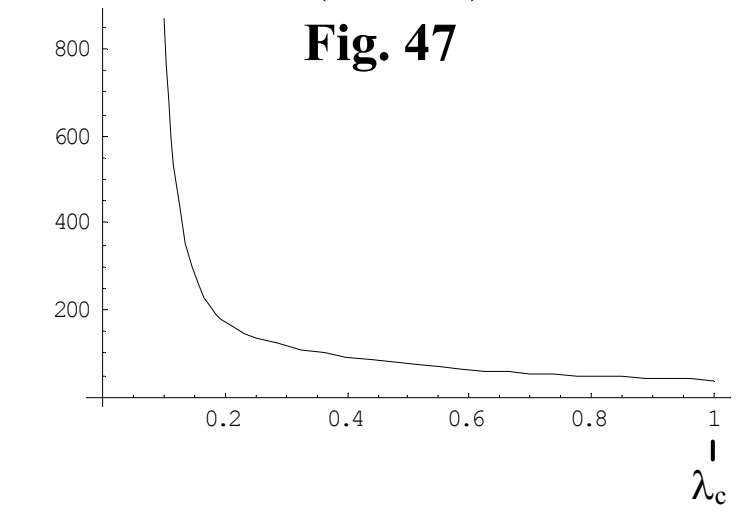

$B_\varphi$ (A vs. r)
**Fig. 48**

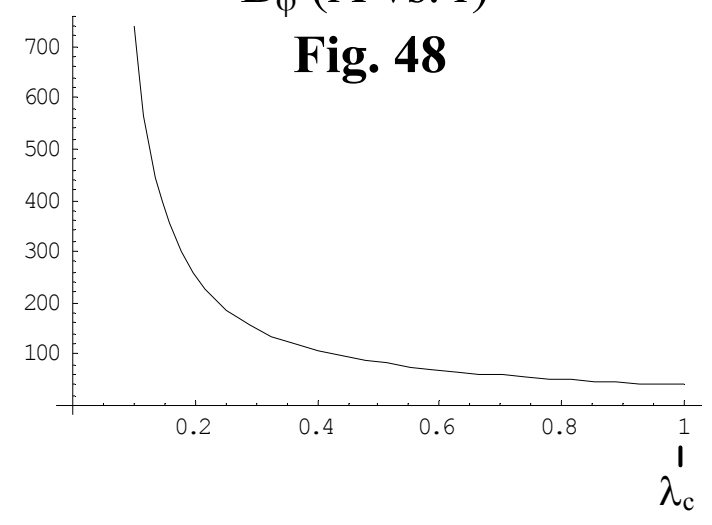

Because of the excellent match between the numerical and theoretical methods, the validity of both methods is confirmed in analyzing the propagation of simple signals in this system. Whereas the theoretical method enables the propagation of simple signals to be clearly understood, the numerical solution is not only useful in verifying the theoretical results, it can also be useful in understanding the propagation of more complex signals which may be difficult to analyze mathematically.

## Experimental verification of $E_\theta$ solution

A simple experimental setup (Fig. 49) has been developed to qualitatively verify the transverse electric field phase vs. distance plot predicted from standard EM theory (Fig. 12) and use it to mathematically determine the phase speed and group speed vs. distance plots (Fig. 13, 14)

**Fig. 49** Experimental setup

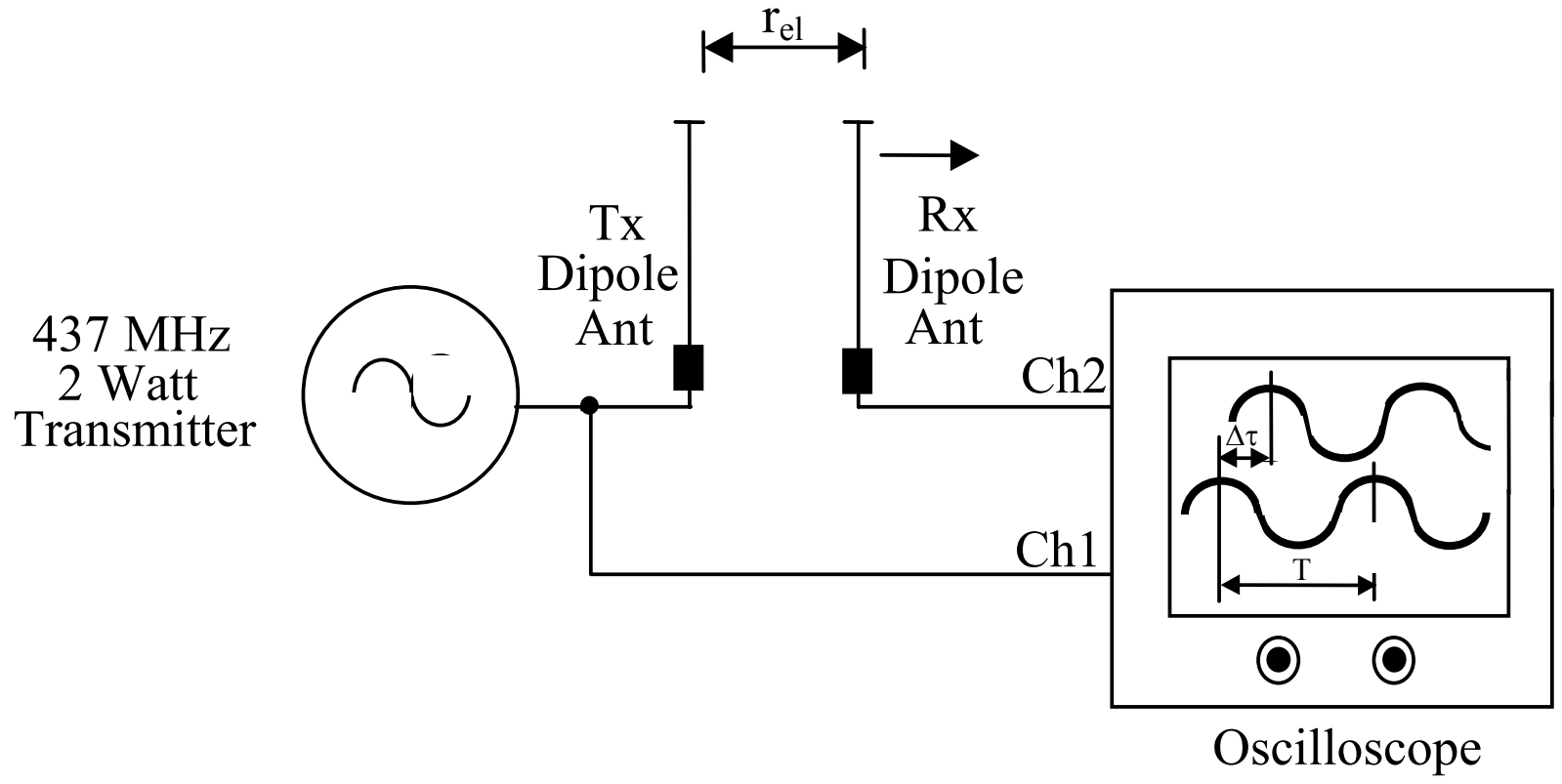

The experiment setup consists of a high frequency UHF FM transmitter (Hamtronics model no. TA451)[1] which generates a 437MHz (68.65cm wavelength), 2 watt sinusoidal electrical signal. The output of the transmitter is connected with a RG58 coaxial cable to a vertical dipole antenna designed for the carrier frequency (model no. RA3126)[2] The output of the transmitter is also connected to channel 1 of the input of a high frequency 500MHz digital oscilloscope (model no. HP54615B). The transmitter output, cable, antenna, and oscilloscope input all have 50 Ohm impedance in order to minimize reflections. A second identical receiver dipole antenna is connected to channel 2 of the high frequency oscilloscope and the antenna is positioned parallel to the vertical transmitting antenna. The sinusoidal signals from the two antennas are monitored with the oscilloscope, triggered to channel 1. The phase difference between the signals is measured using the oscilloscope measurement cursors as the antennas are moved apart from 5 cm to 70 cm in increments of 5 cm (measurements made with a ruler). The oscilloscope calculates the phase from the measured time delay (Δt) and the measured wave period (T): $\theta_{deg} = (360\Delta t)/T$. The phase vs. distance data is analyzed using HPVEE (Ver. 4.01) PC software. The data is then curvefit with a 3rd order polynomial and the data is superimposed to visually verify the accuracy of the curvefit. The phase speed vs. distance curve was then generated by differentiating the resultant curve fit equation with respect to space and using (Eq. 57). The group speed vs. distance curve was generated by using the differential relation: $\Delta\theta = \Delta kr \dfrac{\partial\theta}{\partial kr}$ and inserting it into the relation (Eq. 61): $c_g = \dfrac{\Delta\omega}{\dfrac{\partial\Delta\theta}{\partial r}}$, where: $\Delta\omega = \Delta kc$ yielding:

$$c_g = \frac{360\, c}{\left[ r_{el} \dfrac{\partial^2\theta}{\partial r_{el}^{\;2}} + \dfrac{\partial\theta}{\partial r_{el}} \right]} \qquad \text{where: } r_{el} = \frac{r}{\lambda} \tag{78}$$

[1] Ref. Internet site: www.hamtronics.com
[2] Ref. Internet site: www.elfa.se – part no. 78-069-95

Experimental results

The following graph (ref. Fig. 51) is a plot of the phase vs. distance data (Fig. 50) taken during one experiment. The phase and group speed graphs were generated by curvefitting the experimental data and inserting the curvefit equation into the phase and group speed transformations: (ref. Eq. 57, 78). The first data point is not real and was added to improve the polynomial curvefit. The curvefit yielded the following polynomial:

$$ph = (132.2) + (-262.5)r_{el} + (838.9)r_{el}^2 + (-353.4)r_{el}^3 \quad (79)$$

Experimental data — $E_\theta$ phase plot similar to Fig. 12

**Fig. 50**

| Data # | $r_{el}$ (cm) | Ph (Deg) |
|---|---|---|
| 0 | 0 | 140.0 |
| 1 | 5 | 111.7 |
| 2 | 10 | 102.4 |
| 3 | 15 | 108.6 |
| 4 | 20 | 121.0 |
| 5 | 25 | 136.6 |
| 6 | 30 | 155.2 |
| 7 | 35 | 170.7 |
| 8 | 40 | 195.5 |
| 9 | 45 | 211.0 |
| 10 | 50 | 245.2 |
| 11 | 55 | 282.4 |
| 12 | 60 | 316.6 |
| 13 | 65 | 325.9 |
| 14 | 70 | 366.2 |

**Fig. 51**
θ (deg) vs. r/λ

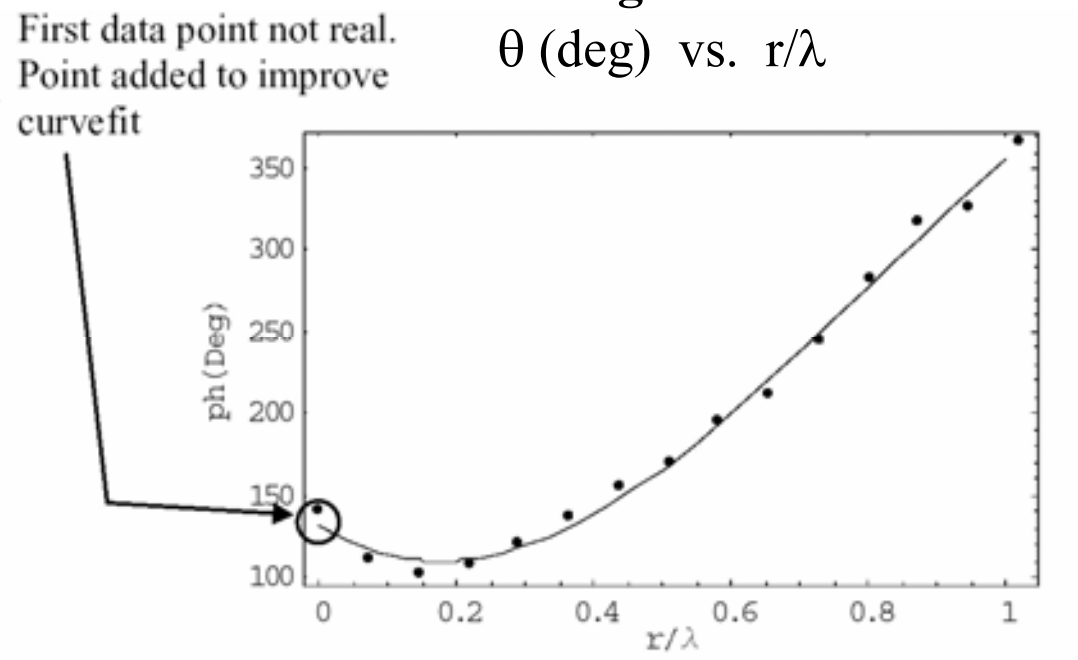

Resultant $E_\theta$ phase speed and group speed plots - very similar to the predicted plots (Fig. 13, 14)

**Fig. 52**
$c_{ph}/c$ vs. r/λ

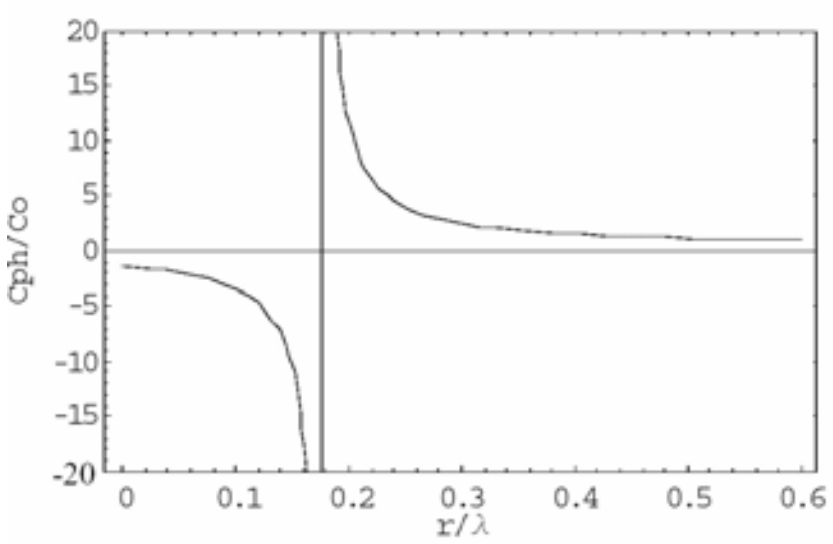

**Fig. 53**
$c_g/c$ vs. r/λ

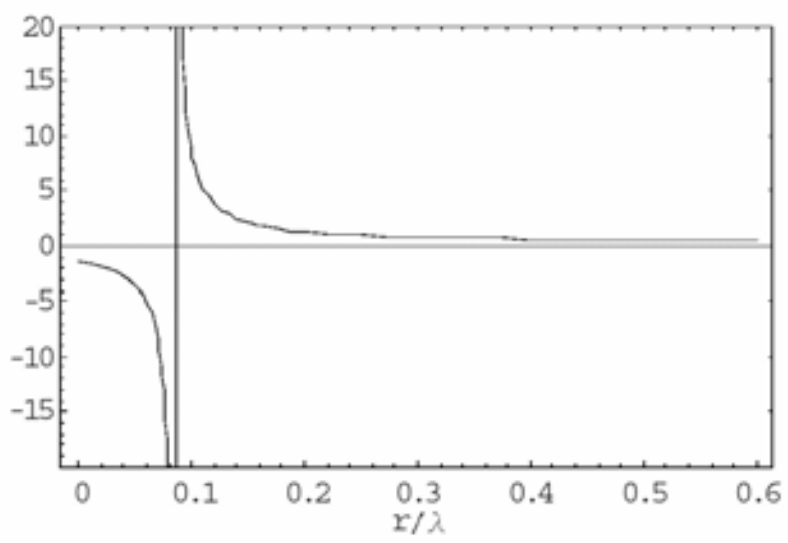

It should be noted that these experimental results are only qualitative due to EM reflections from nearby walls and objects. Quantitative measurements can only be attained in an anechoic chamber. The experiment has been repeated several times in different parts of a 4 x 4m (area) x 2m (height) room at different angular orientations to the walls and the phase vs. distance curve always appears the same within 10%. It is also observed that changing the scope input impedance from 50 Ohms to 1M Ohm input impedance does not noticeably affect the phase vs. distance curve. Since no effect is observed it is concluded that the Tx antenna to Rx antenna variable capacitance combined with the scope input impedance (thereby forming a high pass filter) is not the cause of the phase change. Experimentally it is observed that the electrical field near the source (less than 0.6 λ) is at least an order of magnitude greater than electric field several wavelengths away from the source, which may be reflected. It is concluded that the observed field near the source is predominantly due to near-field effects thereby making the observed results qualitatively reliable. The experimental results (Fig. 51, 52, 53) are qualitatively similar to the electric dipole solution presented (ref. Fig. 12, 13, 14). Differences between experiment and the theory presented can be attributed to EM reflections and also to the fact that the theoretical model for a real dipole antenna is somewhat different from the simple electric dipole solution presented.

## **Other wave propagation systems with similar superluminal behavior**

### Magnetic dipole

Theoretical analysis of a magnetic dipole reveals that the system is governed by the same partial differential equation as the electric dipole with the E and B fields interchanged [20, 12]. The resulting fields are found to be the same as the fields generated by an electric dipole (Eq. 47, 52, 53) and therefore the phase speed and group speed of these fields are the same as (Eq. 67 - 75), except that the E and B fields are interchanged.

### Gravitational quadrapole

It is known that for weak and slowly varying gravitational fields, General Relativity theory reduces to a form of Maxwell's equations. In this limit, Einstein's equation becomes linearized and reduces in MKS units to [21, 22, 9]:

$$\nabla^2 V - \frac{1}{c^2}\frac{\partial^2 V}{\partial t^2} = 4\pi G\rho \qquad (80)$$

Where:

| | |
|---|---|
| $\rho$ = Mass density | $\nabla^2$ = Laplacian |
| V = Gravitational potential | c = Speed of light |
| G = Gravitational constant | t = Time |

Except for the source term, the partial differential equation of the potential is the same as that of an oscillating charge. Because of this similarity one can then use the oscillating charge solutions by simply substituting: $\varepsilon_o = -1/(4\pi G)$. In addition, because momentum is conserved, a moving mass must push off another mass. The gravitational field generated by the secondary mass adds to the gravitational fields generated by the moving mass, resulting in a linear quadrapole source. Although the problem can be solved using only the fields as was done for the dipole in the first part o this paper, simpler methods using the potentials can also be used. One method in particular requires only the scalar potential [23, 3]. The scalar potential for an electric quadrapole is known to be in MKS units [10]:

$$V = -N\left(i\left(\frac{1}{kr} - \frac{3}{(kr)^3}\right) - \frac{3}{(kr)^2}\right)\left(3Cos^2\theta - 1\right)e^{i(kr-\omega t)} \qquad (81)$$

The fields can then be calculated using the following relations:

$$B = \frac{\omega}{c^2}(r \times \nabla V) \qquad E = \frac{ic^2}{\omega}(\nabla \times B) \qquad (82)$$

where (E) is the gravitational force vector and (B) is the solenoidal gravitational force vector. The constant (N) can be determined by substituting the relations: $\varepsilon_o = -1/(4\pi G)$ and q = m into the value of (N) used in the electric quadrapole. In addition, this result can be checked by looking at the static quadrapole solution and comparing it to the above solutions in the limit ($kr \to 0$). The results yield:

$$B_\phi = \frac{6N\omega}{c^2}\left[\left(\frac{1}{kr} - \frac{3}{(kr)^3}\right)i - \frac{3}{(kr)^2}\right]\left[Cos(\theta)Sin(\theta)\right]e^{i(kr-\omega t)} \tag{83}$$

$$E_r = 6Nk\left[\left(\frac{-3}{(kr)^3}\right)i - \frac{1}{(kr)^2} + \frac{3}{(kr)^4}\right]\left[3Cos^2(\theta) - 1\right]e^{i(kr-\omega t)} \tag{84}$$

$$E_\theta = 6Nk\left[\left(\left(\frac{1}{kr}\right) - \frac{6}{(kr)^3}\right)i + \frac{6}{(kr)^4} - \frac{3}{(kr)^2}\right]\left[Cos(\theta)Sin(\theta)\right]e^{i(kr-\omega t)} \tag{85}$$

where: $N = -G\, m\, s^2\, k^3$, G = Grav const., m = mass, s = Dipole length, k = Wave number

The phase and group speed relations for these fields can then be determined by using the phase and group speed equations derived earlier in the paper (Eq. 57, 62). It should be noted that all of the plots look very similar to those of an electric dipole (ref. Fig. 9 - 17).

$B_\phi$ phase, phase speed, group speed analysis

$$ph = kr - ArcTan\left[\frac{kr}{3} - \frac{3}{kr}\right] \underset{kr \ll 1}{\approx} \frac{\pi}{2} + \frac{1}{45}(kr)^5 + O(kr)^7 \tag{86}$$

$$c_{ph} = c_o\left(1 + \frac{3}{(kr)^2} + \frac{9}{(kr)^4}\right) \tag{87}$$

$$c_g = \frac{c_o\left[3(kr)^2 + (kr)^4 + 9\right]^2}{(kr)^4\left[45 + 9(kr)^2 + (kr)^4\right]} \tag{88}$$

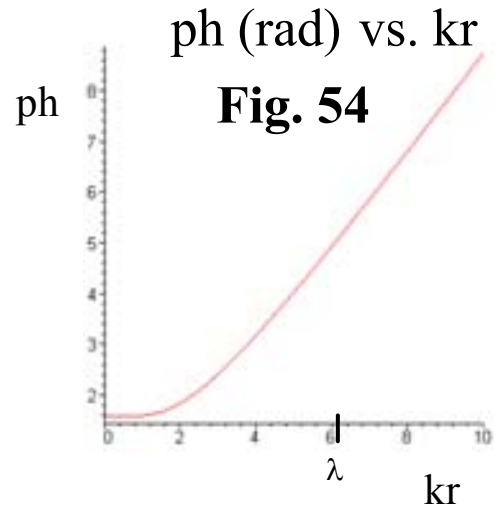

ph (rad) vs. kr
**Fig. 54**

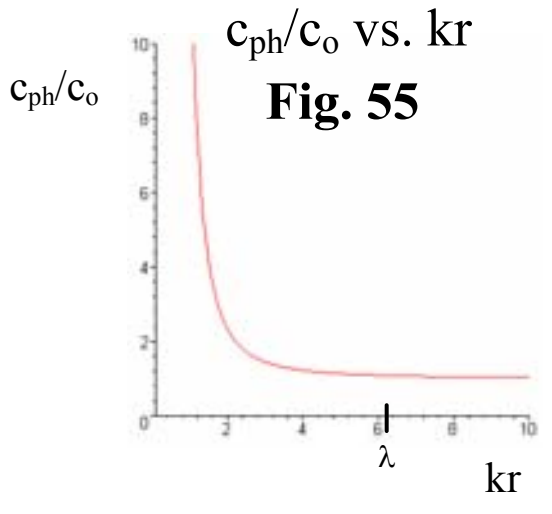

$c_{ph}/c_o$ vs. kr
**Fig. 55**

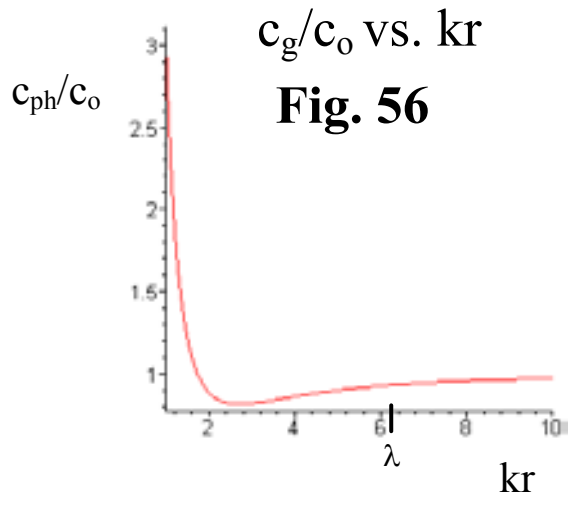

$c_g/c_o$ vs. kr
**Fig. 56**

$E_r$ phase, phase speed, group speed analysis

$$ph = kr + ArcTan\left[\frac{3}{kr - \frac{3}{kr}}\right] \underset{kr \ll 1}{\approx} \frac{1}{45}(kr)^5 + O(kr)^7 \tag{89}$$

$$c_{ph} = c_o\left(1 + \frac{3}{(kr)^2} + \frac{9}{(kr)^4}\right) \tag{90}$$

$$c_g = \frac{c_o\left[(kr)^4 + 3(kr)^2 + 9\right]^2}{(kr)^4\left[(kr)^4 + 9(kr)^2 + 45\right]} \tag{91}$$

ph(rad) vs. kr

**Fig. 57**

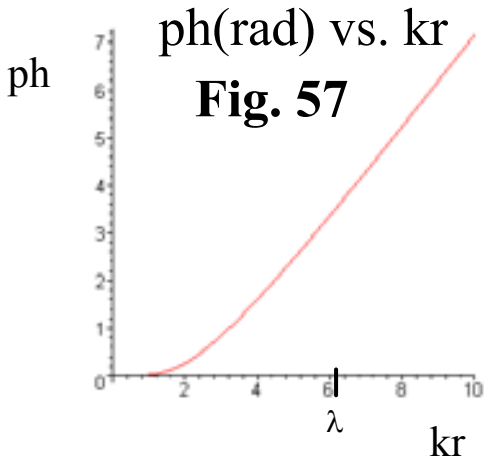

$c_{ph}/c_o$ vs. kr

**Fig. 58**

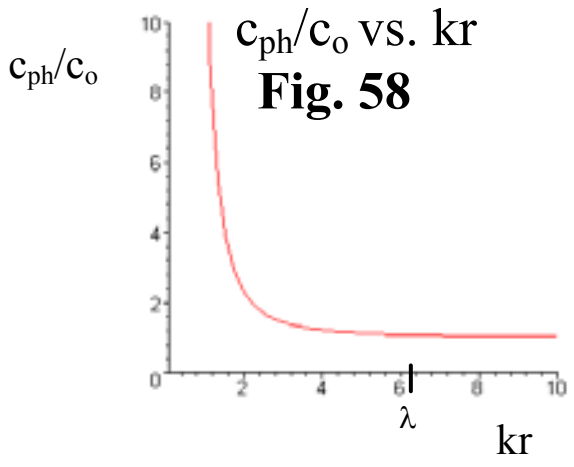

$c_g/c_o$ vs. kr

**Fig. 59**

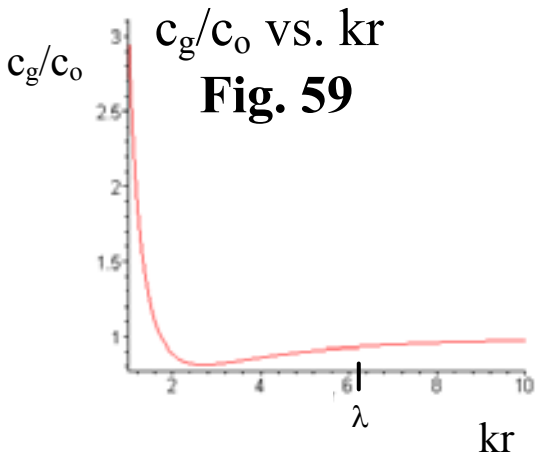

<u>$E_\theta$ phase, phase speed, group speed analysis</u>

$$ph = kr + ArcTan\left[\frac{\frac{-6}{(kr)^3}+\frac{1}{kr}}{\frac{6}{(kr)^4}-\frac{3}{(kr)^2}}\right] \underset{kr \ll 1}{\approx} -\frac{1}{30}(kr)^5 + O(kr)^7 \tag{92}$$

$$c_{ph} = c_o\left(\frac{36-3(kr)^4+(kr)^6}{(kr)^6-6(kr)^4}\right) \tag{93}$$

$$c_g = \frac{c_o\left[36-3(kr)^4+(kr)^6\right]^2}{(kr)^{12}-3(kr)^{10}+18(kr)^8+252(kr)^6-1080(kr)^4} \tag{94}$$

ph(rad) vs. kr

**Fig. 60**

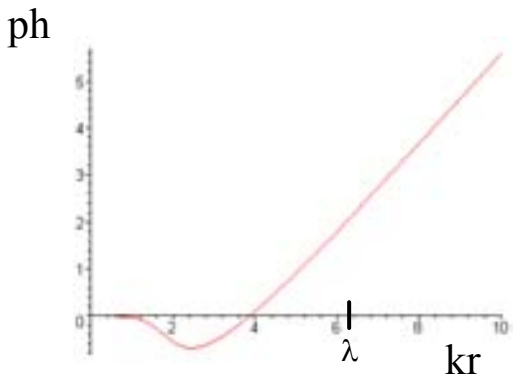

$c_{ph}/c_o$ vs. kr

**Fig. 61**

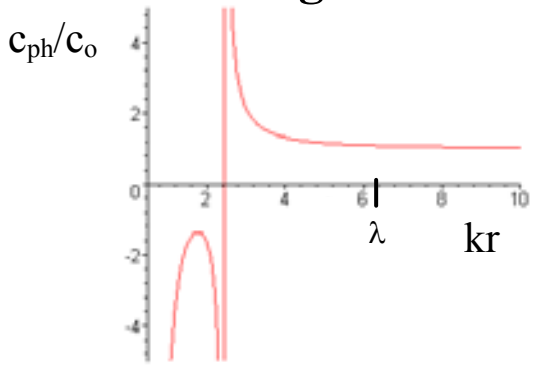

$c_g/c_o$ vs. kr

**Fig. 62**

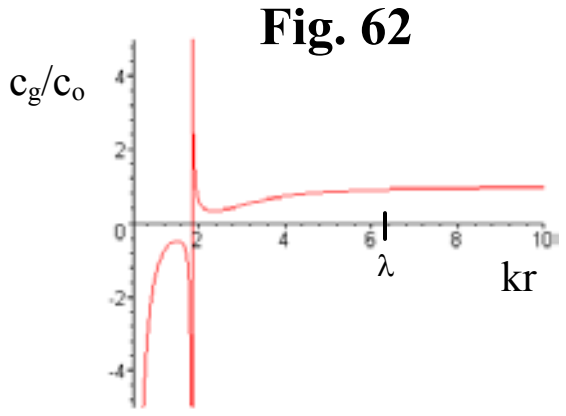

<u>**Field contour plots**</u> (linear quadrapole in center and vertical)

Contour plots of the fields (Eq. 83 - 85) using Mathematica software (ref. Fig. 19) yields:

$B_\phi$ contour plot

**Fig. 63**

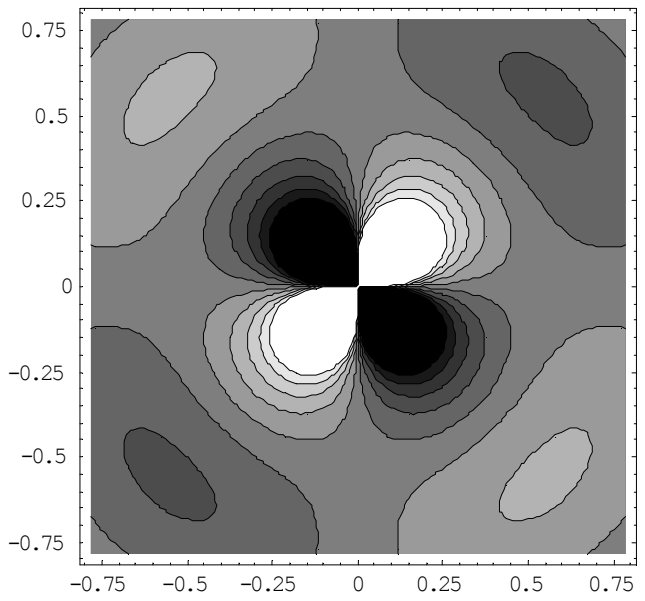

$E_r$ contour plot

**Fig. 64**

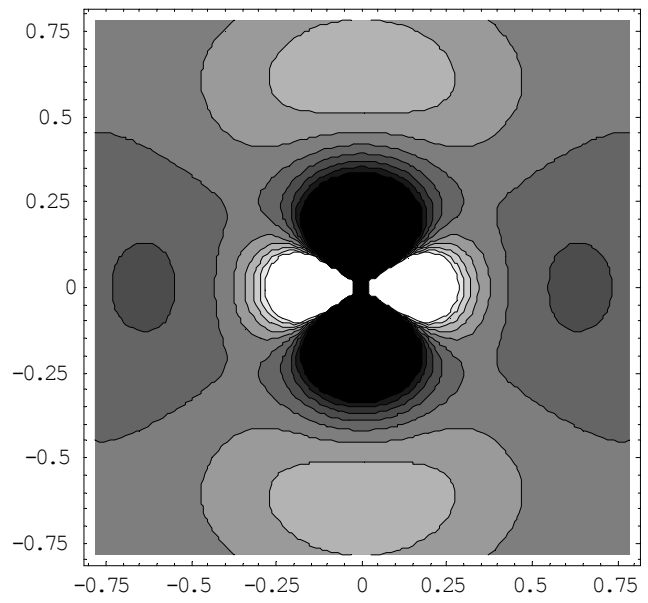

$E_\theta$ contour plot

**Fig. 65**

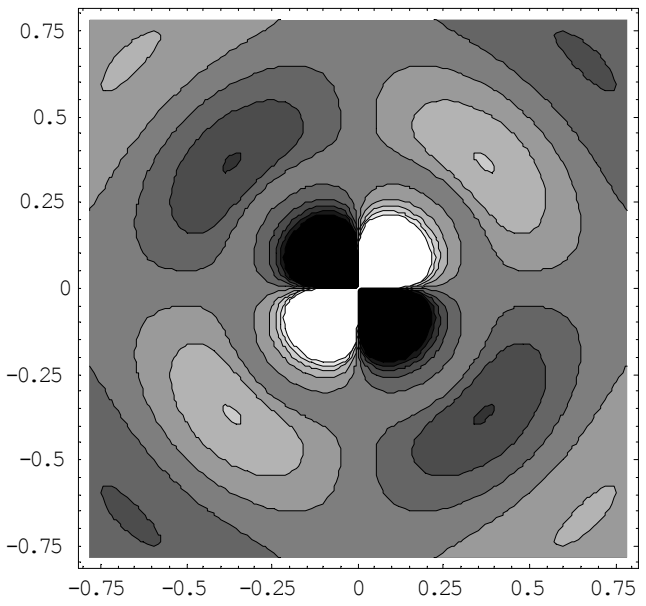

**Total E field plots** (linear quadrapole in center and vertical unless specified)

Using vector field plot graphics in Mathematica software (ref. Fig. 22), the $E_r$ and $E_\theta$ can be combined and plotted as vectors (Fig. 66). A more detailed plot of the total E field can be obtained by using the fact that a line element crossed with the electric field = 0 (ref. Eq. 76, 77). A contour plot (ref. Fig. 24) of the resulting relation yields the total E field plots below (Fig. 67, 68).

Near-field
Vector plot
**Fig. 66**

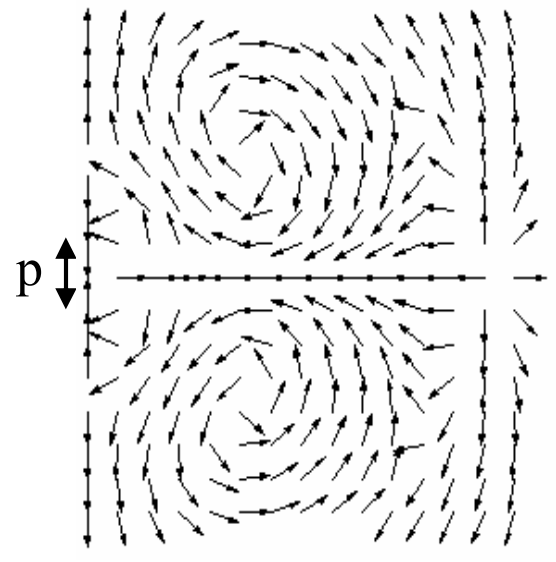

Near-field
Contour plot
**Fig. 67**

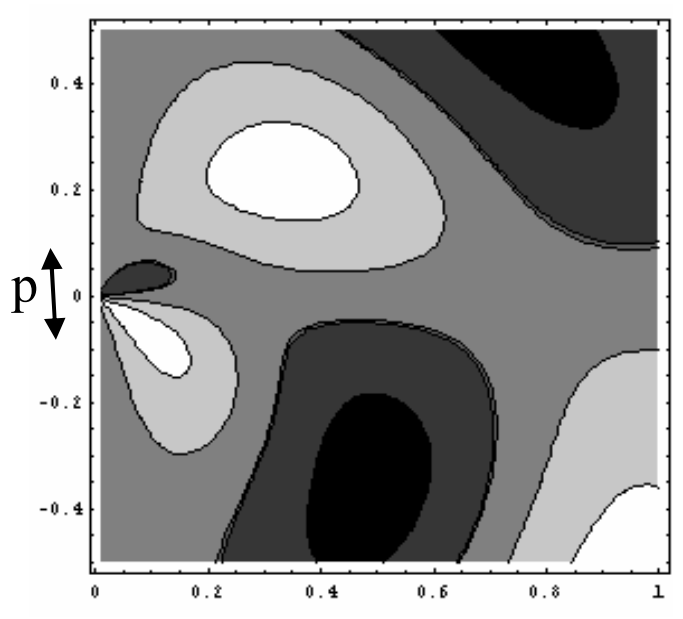

Far-field
Contour plot
**Fig. 68**

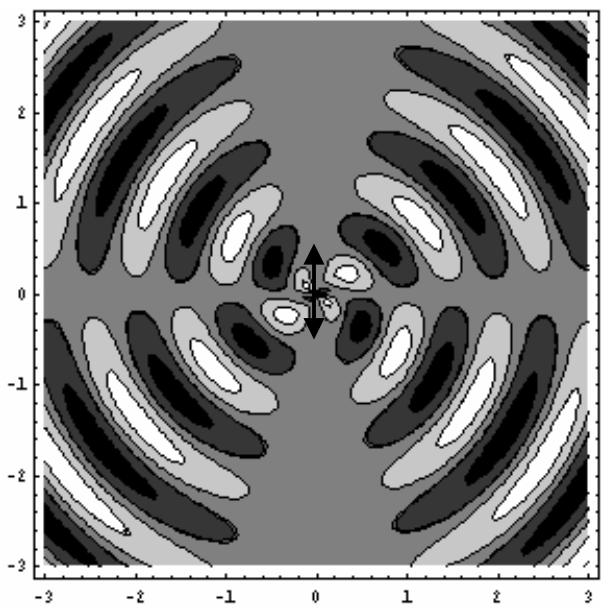

**Experimental evidence of superluminal gravitational fields**

Evidence of near-infinite gravitational phase speed at nearly zero frequency has been observed by a few researchers by noting the high stability of the earth's orbit about the sun [24, 25]. Light from the sun is not observed to be collinear with the sun's gravitational force. Astronomical studies indicate that the earth's acceleration is towards the gravitational center of the sun even though it is moving around the sun, whereas light from the sun is observed to be aberated. If the gravitational force between the sun and the earth were aberated then gravitational forces tangential to the earth's orbit would result, causing the earth to spiral away from the sun, due to conservation of angular momentum. Current astronomical observations estimate the phase speed of gravity to be greater than $2 \times 10^{10}$c. Arguments against the superluminal interpretation have appeared in the literature [26, 27]

**Information speed**

If an amplitude-modulated signal propagates a distance (d) in time (t), then the information contained in the modulation propagates at a speed:

$$c_{inf} = d/(t+T) \qquad (95)$$

where (T) is the amount of time the modulated signal must pass by the detector in order for the information to be determined. The information in the wave is determined by measuring the amplitude, frequency, and phase of the wave modulation envelope.

If a wave is propagated across distances in the farfield of the source, then the wave information speed is approximately the same as the wave group speed. This is because the wave propagation time (t) is much greater than the wave information scanning time (T), consequently:

$$c_{inf} \sim d/t = c_g \tag{96}$$

In the nearfield of the source, if nothing is known about the type of modulation, then the scanning time (T) can be much larger than the wave propagation time (t), thereby making the wave information speed much less than the wave group speed. This can be understood by noting that several modulation cycles are required for a Fourier analyzer to be able to determine the wave modulation amplitude, frequency, and phase. But if the type of modulation is known, then only a few points of the modulated signal need to be sampled by a detector in order to curve fit the signal and therefore determine the modulation information. If the noise in the signal is very small then the signal scanning time (T) can be made much smaller than the signal propagation time (t), consequently: $c_{inf} \sim d/t = c_g$ .

Numerical evidence that information can be extracted in periods much smaller than a modulation wavelength

A numeric simulation was developed using Methematica Ver. 3 software (Fig. 70) to determine if the information contained in an AM signal can be extracted in a period much smaller than a modulation wavelength. To demonstrate this, an AM signal was theoretically specified:

$$AM\ Sig = Cos(\omega_c t)[1 + 7.2\ Cos(\omega_m t)] \tag{97}$$

The signal was then digitized and the resultant data curvefitted, specifying only the carrier frequency ($\omega_c$=10) and the modulation frequency ($\omega_m$=1). The curvefit was able to determine the unknown modulation amplitude given only a few data points in a region much smaller than the modulation wavelength.

**Fig. 69**

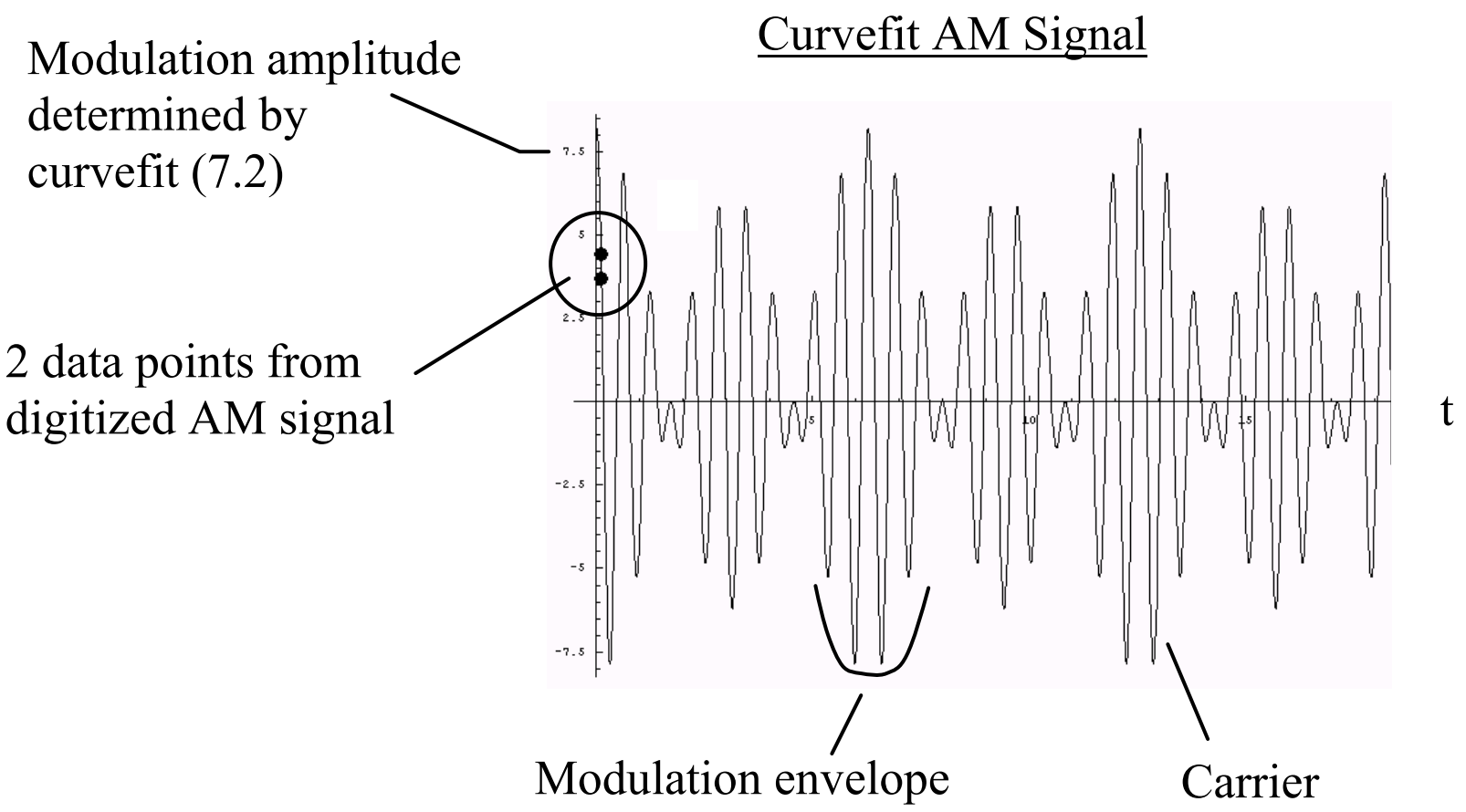

This simulation shows that provided the carrier and modulation frequency are known, a very small part of an AM signal can be curvefit to determine the unknown modulation amplitude (information).

**Fig . 70** Mathematica code for simulation of modulation curvefit

```
(*Program generates an AM signal sampled 0.01 to 0.11 in steps of .01*)
(*Carrier frequency (Wc =10), Modulation frequency (Wm = 1)*)
    Amp=7.2;
    AM=Cos[Wc*t]*(1+Amp*Cos[Wm*t])
    Wc=10; Wm=1;
    Plot[AM,{t,0,20},PlotPoints\[Rule]100]
    points=Table[{t,N[AM]},{t,0.1,0.11,.01}];
    points
    plotpoints=ListPlot[points,PlotStyle->PointSize[0.016]];
    Curvefit=Fit[points,{Cos[Wc*t],Cos[Wc*t]*Cos[Wm*t]},t]
    PlotCurve=Plot[Curvefit,{t,0,20},PlotPoints\[Rule]100];
    Show[plotpoints,PlotCurve];
```

## Relativistic consequences

According to the relativistic Lorentz time transform (Eq. 98), if an information signal can propagate at a speed (w) faster than the speed of light (c), then the signal can be reflected by a moving frame (v) located a distance (L) away and the signal will arrive before the signal was transmitted (Δt′ < 0). Since the information in the signal can be used to prevent the signal from being transmitted, this results in a logical contradiction (violation of causality). How can the signal be detected if it was never transmitted? Consequently, Einstein in 1907 stated that superluminal signal velocities are incompatible with Relativity theory [28].

**Fig. 71**

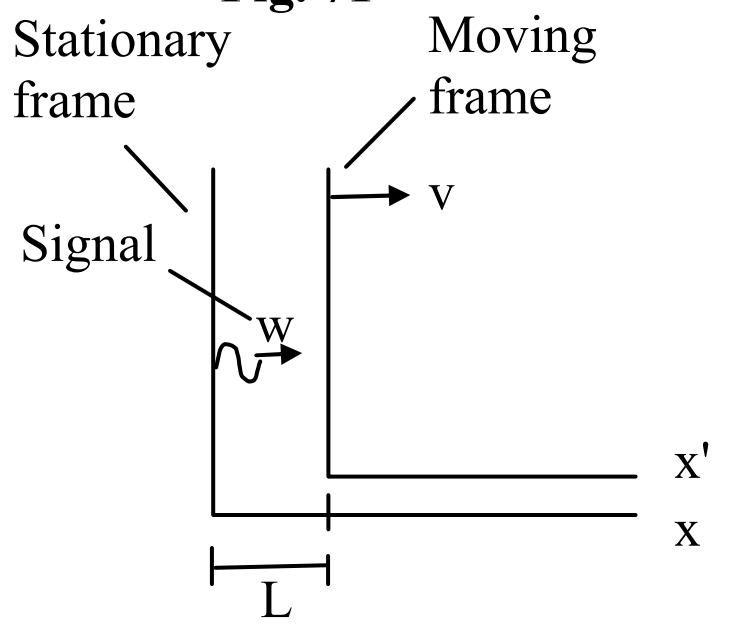

$$\Delta t' = \gamma\left(\Delta t - \frac{v}{c^2}\Delta x\right) \underset{w > \frac{c^2}{v}}{\Rightarrow} \text{Neg} \qquad (98)$$

where: $\Delta t = \dfrac{L}{w}$ $\quad \Delta x = L$

$$\gamma = \frac{1}{\sqrt{1-\frac{v^2}{c^2}}}$$

Because Relativity theory predicts that a moving reflector (which has mass) can never move faster than light (v < c), then in order for (Δt′ > 0) the signal propagation speed must be less than the speed of light (w<c).

## Arguments against the superluminal interpretation

Some physicists have proposed that a dipole source generates position, velocity, and acceleration-dependent propagating fields, each of which propagates at the speed of light [29, 26, 27]. It is argued that the interference of these field components gives the illusion that they propagate superluminally. In the first section of this paper, the electromagnetic fields generated by a dipole source were derived from Maxwell's equations without the use of potentials or gauge equations. The results show the same field solutions as presented by other authors (Eq. 47, 52, 53) but differ in that the electric field is shown to be generated only by the position of the dipole and not the combination of the dipole's position, velocity and acceleration (Eq. 52, 53). Similarly the magnetic field is shown to be generated only by the velocity of the dipole and not the combination of the dipole's, velocity and acceleration (Eq. 47). It is shown that the real and imaginary components of the fields are due to the Taylor expansions of the spatial part of the fields (Eq. 40) and not due to time derivatives of the source as suggested by other authors.

Another common argument mentioned by some physicists is that a source impulse yields an impulse propagating at the speed of light. Referring to the expression for the retarded potentials (Eq. 35), one can see that the potentials are expressed in terms of source position (R). But conversion to the origin coordinate (r) yields a very different result, which is dispersive in the nearfield (ref. Eq. 43 with higher order terms in $\xi$ included). In addition the fields, which are what are measurable yield even more dispersive results (ref. Eq. 49, 56, 57), where the effects are observed to the zeroth order in $\xi$. This indicates that an impulse will distort as it propagates in the nearfield, and reduces to the speed of light in the farfield since the higher order terms decay more rapidly in the farfield [ref. argument corresponding to (Fig. 38), and also (ref. Fig. 9, 12, 15) showing: $\theta = f(\omega)$ for given r].

Some physics accept that phase velocity and group velocity in these systems can be superluminal but that the information speed is less than the speed of light. It has been shown in this paper that although group speed can differ from information speed, provided the noise is small and the method of modulation is known, group speed can be approximately the same as the information speed (ref. section corresponding to Eqn. 95 - 97). It is also commonly stated by physicists that the front speed (speed of a field step function or impulse) is limited to the speed of light. In the above paragraph it has been argued that an impulse changes shape as it propagates, and therefore it can not be used to determine the speed of the field in the nearfield. The analysis in this paper has shown that in order for signals to propagate without much dispersion, the signals must be narrowband, such as provided by conventional AM, FM, PM modulations (ref. Group speed analysis section corresponding to Eq. 62 - 66). This is because the phase vs. frequency curve must be linear over the bandwidth of the signal. Because impulses and step functions are broadband signals, different frequency components will propagate at different speeds resulting in signal distortion.

## Conclusion

The analysis presented in this paper has shown that the fields generated by an electric or magnetic dipole, and also the gravitational fields generated by a quadrapole mass source, propagate superluminally in the nearfield of the source and reduce to the speed of light as they propagate into the farfield. The group speed of the waves produced by these systems has also been shown to be superluminal in the nearfield. Although information speed can be less than group speed in the nearfield, it has been shown that if the method of modulation is known and provided the noise of the signal is small enough, the information can be extracted in a time period much smaller than the wave propagation time. This would therefore result in information speeds only slightly less than the group speed which has been shown to be superluminal in the nearfield of the source. It has also been shown that Relativity theory predicts that if an information signal can be propagated superluminally, then it can be reflected by a moving frame and arrive at the source before the information was transmitted, thereby enabling causality to be violated.

Given these results, it is at present unclear how to resolve this dilemma. Relativity theory could be incorrect, or perhaps it is correct and information can be sent backwards in time. Perhaps as suggested by the 'Hawking chronology protection conjecture' [30], nature will intervene in any attempt to use the information to change the past. Therefore information can be propagated backwards in time but it cannot be used to change the past, thereby preserving causality. Another possibility is that according to the 'many-worlds' interpretation of quantum mechanics [31], multiple universes are created any time an event with several possible outcomes takes place. If this interpretation is correct, then information can be transmitted into the past of alternative universes, thereby preserving the past of the universe from which the signal was transmitted.

In addition to the theoretical implications of the research discussed above, it may also have practical applications, such as increasing the speed of electronic systems that will soon be limited by light-speed-time delays. It should also be possible to reduce the time delays inherent in current astronomical observations by monitoring lower frequency EM and eventually gravitational radiation from these sources. Lastly, using low frequency EM transmissions, it should be possible to reduce the long communication time delays to spacecraft.